\documentclass[trackchanges,twocolumn]{aastex701}
\usepackage{graphicx}
\usepackage{xcolor}
\usepackage{keyval}
\usepackage{hyperref}
\usepackage{graphicx}	
\usepackage{amsmath}	
\usepackage{dirtytalk}  
\usepackage{float}
\def\heii{He\,{\sc ii}~}
\def\nv{N\,{\sc v}~}

\def\oiii{[O\,{\sc iii}]\,}
\def\oii{[O\,{\sc ii}]\,}

\def\civ{C\,{\sc iv}~}

\def\feii{Fe\,{\sc ii}~}

\def\caii{Ca\,{\sc ii}~}

\def\kms{km~s$^{-1}$}

\def\reds{red NLSy1s~}
\def\blues{blue NLSy1s~}
\def\ldnls{Intermediate Seyferts~}
\def\ebv{$E(B-V)$~}

\def\hb{H$\beta$~}
\def\halpha{H$\alpha$~}

\begin{document}

\title{Anomalous narrow line Seyfert I galaxies from SDSS DR17}

\author[orcid=0009-0007-5545-0518,sname='Tiwari']{Arihant Tiwari}
\affiliation{Indian Institute of Astrophysics, 100 Feet Rd, Santhosapuram, 2nd Block, Koramangala, Bengaluru, Karnataka, India}
\email[show]{tivariarihant@gmail.com}  

\author[orcid=0000-0003-4592-447X,sname=' ']{Rachana} 
\affiliation{Indian Institute of Astrophysics, 100 Feet Rd, Santhosapuram, 2nd Block, Koramangala, Bengaluru, Karnataka, India}
\affiliation{Joint Astronomy Programme, Department of Physics, Indian Institute of Science, Bengaluru, 560012, India}
\email{rachana2022@iisc.ac.in}

\author[0000-0001-5937-331X,sname='Vivek']{M. Vivek}
\affiliation{Indian Institute of Astrophysics, 100 Feet Rd, Santhosapuram, 2nd Block, Koramangala, Bengaluru, Karnataka, India}
\email{vivek.m@iiap.res.in}

\author[0000-0002-8377-9667,sname='Rakshit']{Suvendu Rakshit}
\affiliation{Aryabhatta Research Institute of Observational Sciences, Nainital–263001, Uttarakhand, India}
\email{suvenduat@gmail.com}

\begin{abstract}

We present an analysis of 22,656 narrow-line Seyfert 1 galaxies (NLSy1s) from Sloan Digital Sky Survey (SDSS) DR17 ($0.1\leq z\leq 0.9$), identifying a sample of spectroscopically anomalous sources. These anomalies were detected via the spectroscopic quasar anomaly detection (\texttt{SQuAD}) algorithm, which employed principal component analysis and hierarchical k-means clustering. Various physical diagnostic analysis were performed such as the color excess ($E_{(B-V)}$) calculations, Wide-field Infrared Survey Explorer (WISE) color analysis, probing \oiii equivalent width as an inclination indicator, the Baldwin, Phillips \& Terlevich (BPT) diagram and eigenvector 1 diagram. We detected 620 anomalous NLSy1s classified into two groups i.e. 246, \reds, exhibiting host galaxy dominated spectra with a low luminosity active galactic nuclei (AGN) core revealed by the emission line widths. Another set of 374 Blue NLSy1s, strongly luminous galaxies with enhanced AGN activity, bluer continuum as compared to a typical NLSy1 and stronger \feii emission. Finally, the third group of 257 outliers, identified as Intermediate Seyferts, a class of Seyfert galaxies identified by composite emission profiles, and extremely strong emission lines paired with virtually no continuum. These sources also exhibit rare and high ionization emission lines unseen in any other NLSy1 spectra (e.g. [Ne\,{\sc v}]$\lambda3345$, Ne\,{\sc v}$\lambda3426$, Ne\,{\sc iii}$\lambda3869$ etc). We conclude that the differentiating factor between red and \blues is not dust obscuration or orientation effect, but intrinsic distinction in AGN activity. The resulting sample is presented as a value-added catalog.

\end{abstract}




\section{Introduction}

Active galactic nuclei (AGNs) are among the most energetic and dynamic objects in the universe, powered by the accretion of matter onto supermassive black holes (SMBHs) located at the centers of galaxies \citep[][]{shlosman1990fuelling}. These luminous sources display a wide range of observational characteristics, leading to their classification into various subtypes based on their spectral features, luminosity, and variability. Quasars, the most luminous class of AGNs, typically display both broad and narrow emission lines, along with a continuum that follows a power-law, $F(\nu) \propto \nu ^{- \alpha}$, with a spectral index of approximately $0.44$ between 1300 and 5000~\AA. This index steepens to about $2.45$ at wavelengths beyond $\sim$5000~\AA, a spectral break that is primarily attributed to increasing host galaxy contamination at lower redshifts \citep{2001AJ....122..549V}. Moreover, the evolution of AGNs follows the quasar luminosity function (QLF), which peaks at redshifts $z \sim 2$–3, tracing the peak era of black hole growth in the universe. This evolution provides crucial insights into the role of quasars in cosmic feedback processes and the co-evolution of galaxies and their central black holes \citep{yu2002observational}.

Among these various AGN subclasses, narrow-line Seyfert 1 galaxies (NLSy1s) represent a particularly distinct and intriguing population. Spectroscopically, they are characterized by a relatively narrow \say{broad} component of the H$\beta$ emission line (full width at half maximum, FWHM $<$ 2000~\kms), weak \oiii emission with a flux ratio of \oiii /H$\beta \leq 3$, and strong \feii emission, typically with \feii/H$\beta_{\mathrm{tot}} > 0.5$ \citep[][]{Osterbrock1985,goodrich1989spectropolarimetry, 2001A&A...372..730V}. The prominent \feii emission in AGNs is often found to be anti-correlated with both the strength of \oiii and the width of the broad Balmer lines, suggesting a complex interplay between the ionized gas and the central engine \citep{1992ApJS...80..109B}. The distinct observational characteristics of NLSy1s are believed to stem from the unique physical conditions in their central engines. Black hole masses in NLSy1 galaxies are typically estimated using the FWHM of the broad H$\beta$ line and the continuum luminosity via single-epoch methods calibrated against reverberation mapping \citep{kaspi2000reverberation, kaspi2005relationship,peterson2015measuring}. In these sources, the broad H$\beta$ profiles are generally better represented by Lorentzian rather than Gaussian shapes. The second-order moment, often used for Gaussian profiles \citep{dalla2020sloan}, is therefore not suitable in this case. These estimates suggest that NLSy1s host relatively low-mass SMBHs, generally in the range of $10^6$ to $10^8$ solar masses \citep{2011arXiv1109.4181P}, and accrete at high rates, often close to the Eddington limit \citep{1995MNRAS.277L...5P}. 

% , making FWHM, a more reliable measure of the line width.

Due to their low black hole masses and high accretion rates, NLSy1 galaxies exhibit some of the most extreme multiwavelength signatures among Type 1 AGNs, particularly in the X-ray and UV regimes. In the X-ray regime, NLSy1s frequently exhibit steep soft X-ray spectra (E $<$ 2 keV), a pronounced soft excess, and rapid, large amplitude variability, often exceeding 40$\%$, on timescales ranging from hours to years \citep{boller1996soft, 1999ApJS..125..297L, 1999ApJS..125..317L, 2004AJ....127.1799G}. These properties are typically more extreme than those observed in their broad-line Seyfert 1 (BLSy1) counterparts. Spectral complexity is further enhanced by signatures of ionized winds and outflows \citep{2002ApJ...565...78B, 2005A&A...431..111B}. Multiwavelength monitoring has revealed correlated variability across X-ray, UV, and optical bands, supporting a scenario where fluctuations in the inner accretion flow modulate reprocessed emission at longer wavelengths \citep[e.g.][]{2010ApJS..187...64G}. While some NLSy1s are X-ray weak, emitting less X-ray flux relative to their optical/UV output, others, especially radio-loud or jet-dominated sources, exhibit strong non-thermal X-ray emission from synchrotron or inverse Compton processes \citep{2015MNRAS.453.4037O,2023MNRAS.523..441Y, 2025arXiv250404492C}.

The presence of strong non-thermal X-ray emission in some NLSy1s is closely tied to their radio and high-energy behavior. A subset of NLSy1s, classified as radio-loud, exhibit compact radio cores and flat or inverted spectra, with some sources showing signs of relativistic jet activity such as high brightness temperatures or apparent superluminal motion. In several cases, they have also been detected at $\gamma$-ray energies by the Fermi gamma-ray space telescope \citep{Abdo2009, 2011nlsg.confE....F, 2018MNRAS.477.5127Y}, confirming the presence of powerful jets. Intriguingly, some of these jet-dominated NLSy1s host relatively massive black holes, with estimates reaching $\sim10^7$–$10^8~M_\odot$, i.e. at the upper end of the typical NLSy1 range ($10^6$–$10^8~M_\odot$; \citealt{2011nlsg.confE....F}), and comparable to those of blazars and BLSy1s \citep{2013MNRAS.431..210C, 2016MNRAS.458L..69B}. Their broadband spectral energy distributions (SEDs) often resemble those of blazars, with synchrotron and inverse Compton peaks spanning the radio to $\gamma$-ray regime. However, \citet{2020Univ....6..136F} pointed out that these analyses have several limitations and suggested that some jetted NLSy1s might still host relatively low-mass black holes. In the ultraviolet regime, NLSy1 galaxies display distinctive spectral features that further underscore their complexity. Many NLSy1s exhibit redder UV continua compared to typical quasars, often attributed to intrinsic dust extinction or line-of-sight absorbers \citep{2003PASP..115..592C}. Broad, blueshifted absorption lines in high-ionization species such as \civ and \nv have been observed in several cases, signaling the presence of radiation-driven disk winds or warm absorbers (WAs) \citep[e.g.][]{2004ApJ...611..107L, 2004ApJ...611..125L, 2013MNRAS.430.1102T}. Highly ionized X-ray ultra-fast outflows have also been detected in some NLSy1s \citep{2023ApJ...952...52R}. These features provide valuable diagnostics of the coupling between the accretion disk and the outflowing gas, and, when combined with X-ray warm absorber signatures, offer a multi-phase picture of AGN feedback in NLSy1s.

% Intriguingly, some of these jet-dominated NLSy1s host black holes with estimated masses $10^7$–$10^8~M_\odot$, similar to their counterparts BLSy1.
% rather than ultra-fast outflows
NLSy1 galaxies have also been extensively studied in the infrared band. AGN emission in this regime is primarily driven by thermal radiation from the dusty circumnuclear torus, which absorbs high-energy ultraviolet and optical photons from the accretion disk and re-emits them at longer wavelengths. This reprocessed mid-IR emission serves as a powerful tracer of nuclear activity, particularly in obscured systems where optical and UV light are significantly attenuated by dust. Using data from the Wide-field Infrared Survey Explorer (WISE), \citet{2019MNRAS.483.2362R} analyzed 492 NLSy1 galaxies and reported a mean $W1-W2$ color of 0.99 ± 0.18 mag. For general AGN selection, \citet{stern2012mid} and \citet{assef2013mid} proposed $W1-W2$ thresholds of $\geq 0.8$ and $\geq 0.662$, respectively, as robust criteria to distinguish AGNs from inactive galaxies, since redder $W1{-}W2$ colors trace stronger hot-dust emission from the torus. \citet{mingo2016mixr} further refined these diagnostics by combining $W1-W2$ and $W2-W3$ color cuts, identifying star-forming Seyfert galaxies with $W1-W2 > 0.5$ and $0.6 < W2-W3 < 3.4$. These red mid-IR colors reflect strong thermal emission from the AGN-heated torus, while lower values typically indicate increased host galaxy contamination. As a result, mid-infrared color selection remains effective for uncovering both unobscured (Type 1) and heavily obscured (Type 2) AGNs, including those often missed in optical and X-ray surveys due to dust extinction or weak nuclear signatures.

Due to the diversity of emission mechanisms and complex physical conditions in their nuclear regions, some NLSy1s exhibit unusual or anomalous spectroscopic behavior. For example, \citet{zhang2022discovery} discovered a late-time X-ray flare and anomalous emission line enhancement after the nuclear optical outburst. \citet{steinhardt2013quasars} present a sample of quasars with anomalous \hb profiles, featuring several NLSy1s. \citet{nagar2002ngc} presented strong evidence that NGC 5506 is an NLSy1, despite previously being classified as an intermediate or type 2 Seyfert. This galaxy displays several anomalous properties, including being optically obscured, hosting both a type 1 AGN and a nuclear water vapor megamaser, and having a relatively high black hole mass for an NLSy1. These examples underscore the need for systematic identification of outliers within the NLSy1 population that can provide in situ conditions to study certain properties of NLSy1s in isolation for a deeper understanding.

This project leverages the spectroscopic quasar anomaly detection algorithm \citep[\texttt{SQuAD};][]{tiwarisquad1} to identify anomalous NLSy1 galaxies based on their spectroscopic features and to perform detailed follow-up analyses of the most intriguing cases. This work represents the second installment in a series of studies aimed at systematically uncovering and characterizing spectroscopic outliers in AGN populations. In this paper, we focus specifically on anomalies within the rest-frame optical spectra of a curated sample of NLSy1 galaxies drawn from Sloan digital sky survey data release 17 \citep[SDSS-DR17;][]{2022ApJS..259...35A}, as compiled by \citet{paliya2024narrow}. \citet{paliya2024narrow} classified the NLSy1s solely on the basis of the definition markers i.e. \hb FWHM and \oiii to \hb ratio, making it possible for contaminants such as intermediate Seyferts and quasars to be included in the catalog. We expect to utilize the efficiency of the \texttt{SQuAD} algorithm in objectively picking out abnormal spectra to extract such sources too along with the any anomalous NLSy1s. The paper is organized as follows: In Sect.\ref{Methodology}, we describe the sample selection, and the application of \texttt{SQuAD} algorithm to extract the anomalous NLSy1s. Sect.\ref{Results} presents our main findings, with visual representation of the detected anomalies and their observed spectral features. In Sect.\ref{sec: Physical Diagnostics} we discuss various physical diagnostic procedures utilized to reveal the physical properties of the anomalies and their implications. Finally, Sect.\ref{sec: Discussion} summarizes the findings and traces out the follow-up study plans.

\section{Methodology} \label{Methodology}

\subsection{Data}

The galaxies analyzed in this project are obtained from the \citet{paliya2024narrow} catalog which contains 22,656 sources identified as NLSy1s by carrying out a detailed decomposition of optical spectra of quasars and galaxies from SDSS DR17. The sample contains sources in the redshift range $z\in[0.13,0.90]$, the spectra for which were obtained using the
SDSS and baryon acoustic oscillation spectroscopic survey
(BOSS) spectrographs from the SDSS survey, covering wavelengths from 3600 to 10,400 \AA\,(with good throughput between 3650 and 9500 \AA) and a resolution of 1560–2270 in the blue channel and 1850–2650 in the red channel \citep[][]{smee2013multi}. The redshift range translates to a common wavelength window of 3600 to 5400 \AA, capturing six prominent emission lines in a typical NLSy1 spectrum: \oiii$\lambda5007,4959$\AA, H$\beta\,\lambda4861$\AA, H$\gamma\,\lambda4340$\AA, H$\delta\,\lambda4102$\AA, {[Ne\,{\sc iii}]}$\lambda3869$\AA,\, {[O\,{\sc ii}]}$\lambda3727$\AA\,\citep[e.g.][]{schmidt2016spectral,constantin2003ultraviolet}.\\
The spectra are analyzed as per the \texttt{SQuAD} algorithm to detect, group, and characterize the anomalous NLSy1 galaxies. The following steps were performed chronologically to obtain the anomalous objects.

\begin{figure*}
    \centering
    \includegraphics[width=1\linewidth]{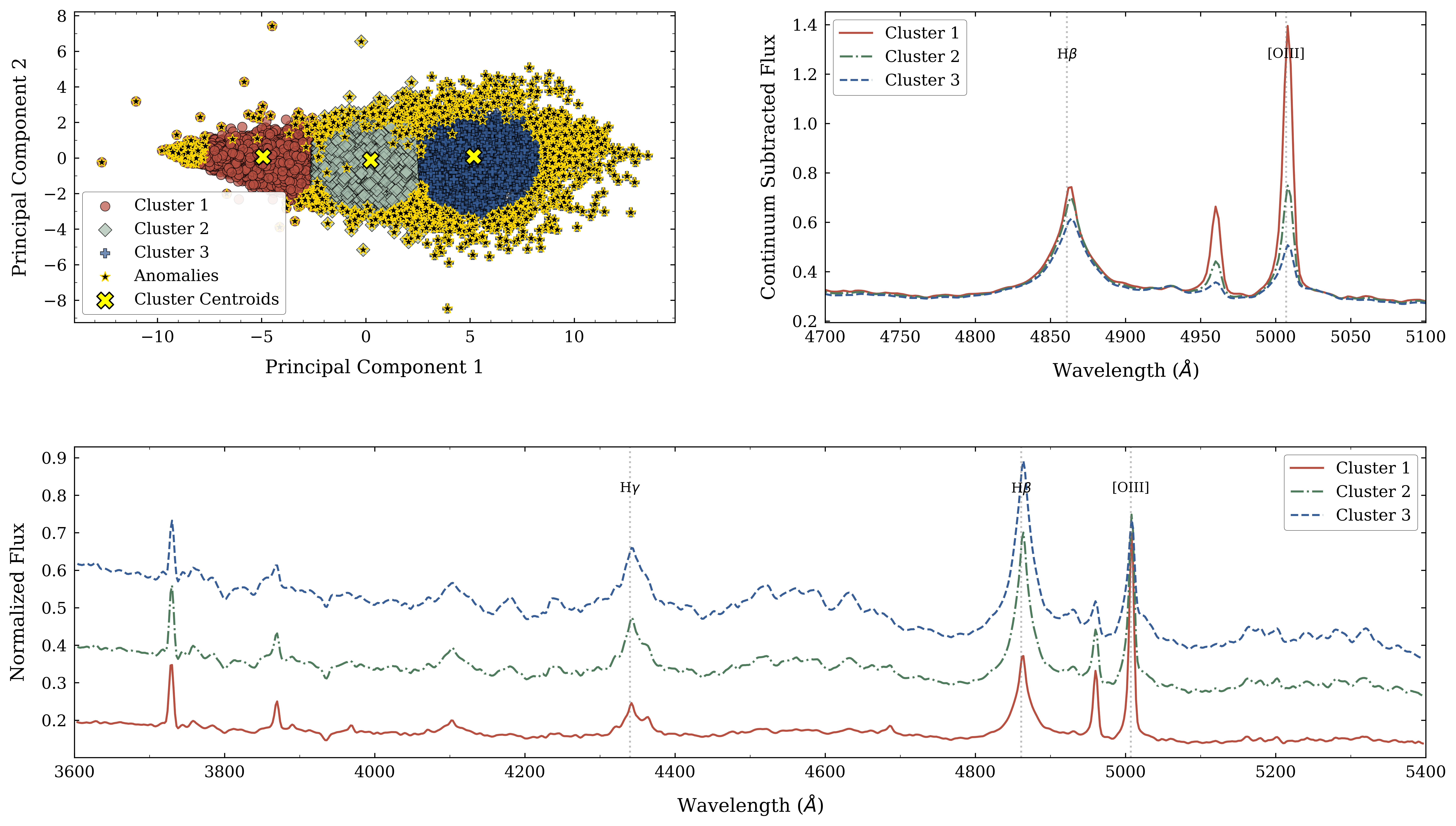}
    \caption{Top Left: A 2D projection (PCA 1 versus PCA 2 coefficients) of the datasets. The dataset is divided into three clusters (cluster 1: brown, cluster 2: green, cluster 3: blue) within the 20-dimensional PCA
    hyperspace, using k-means clustering. Sources classified as anomalous after applying a $Q3 + 1.5 \times IQR$ threshold are shown as golden scatter points overlayed on top of the cluster members. The cluster centroids are marked by yellow crosses. Top Right: Continuum subtracted mean composite spectrum for each cluster to highlight the differences in emission line strength. Bottom: Mean composite spectrum for each cluster of the dataset. The color of each spectrum corresponds to the color of the cluster as shown in the left panel.}
    \label{fig:Cluster and Cluster Mean Spectra}
\end{figure*}

\subsection{Spectroscopic quasar anomaly detection (SQuAD)} \label{subsection: SQuAD Algorithm}
During this work, we followed the \texttt{SQuAD} algorithm exactly, with no additional modifications. Details of the algorithm can be found in \cite{tiwarisquad1}. We briefly summarize the algorithm below as applied here.  
\begin{enumerate}
    \item Pre-processing: The raw spectra obtained from SDSS are re-sampled with a 2\AA\, binning to bring all the spectra to the same wavelength grid and simultaneously reduce the size of the spectral array by a factor of two. These resampled spectra are smoothed using a Savitzky-Golay filter \citep{1964AnaCh..36.1627S} which helps reduce the noise and unwanted artifacts. Once smoothed, all spectra are max-normalized to bring the flux range to [-1,1] to ensure uniformity, needed for the anomaly detection algorithm. These processed spectra are then fed into the anomaly detection pipeline.

    \item Principal component analysis: We apply principal component analysis \citep[PCA;][]{hotelling1933analysis} with 20 components to reduce the size of each spectrum from a 900 dimensional wavelength-flux vector to 20 dimensional vector in the PCA hyperspace. The number of components was chosen as to achieve more than 95\% cumulative explained variance, which was achieved with 20 components accounting for 95.2\% total explained variance. 

    \item Clustering: We then employed k-means clustering \citep{macqueen1967some} to categorize the NLSy1 galaxies based on the Euclidean separation between the PCA eigenvalues. K-means clustering requires a pre-defined number of clusters $(k)$ into which the dataset is supposed to be divided. We calculated the optimum number of clusters using elbow method on the sum of squared errors (SSE) and silhouette coefficients. Based on this $k=3$ was obtained as the optimum number of clusters for our dataset.

    \item Anomaly detection: Once clustered, we computed the Euclidean distance of each data point from its respective cluster centroid and analyzed the resulting distribution. Statistical tests confirmed a non-normal, long-tailed spread, best described by a Gumbel distribution \citep{gumbel1935valeurs}. Given this, we used the interquartile range \citep[IQR;][]{tukey1977exploratory} method to identify outliers, marking data points beyond \( Q3 + 1.5 \times IQR \) as anomalies. This resulted in 944 outliers, which we refer to as \say{anomalous galaxies} here onward. \label{subsection:Anomaly Detection}

    \item Anomaly grouping: On visual inspection, it was observed that the detected anomalies exhibited repeating trends. Hence, to further categorize these anomalies, we re-applied k-means clustering solely on the outliers or anomalies detected in the previous step. Using a 20-component PCA, which captured ~97\% of the variance, and the elbow method, we determined an optimal cluster number of three $(k=3)$. This step ensured that similar spectral anomalies were grouped together, independent of their original cluster assignment. \label{subsec: Anomaly Grouping}
    
\end{enumerate}
A detailed flowchart of the \texttt{SQuAD} algorithm can be found in the Figure A.1 in the appendix of \cite{tiwarisquad1}.

\begin{figure*}
    \centering
    \includegraphics[width=1\linewidth]{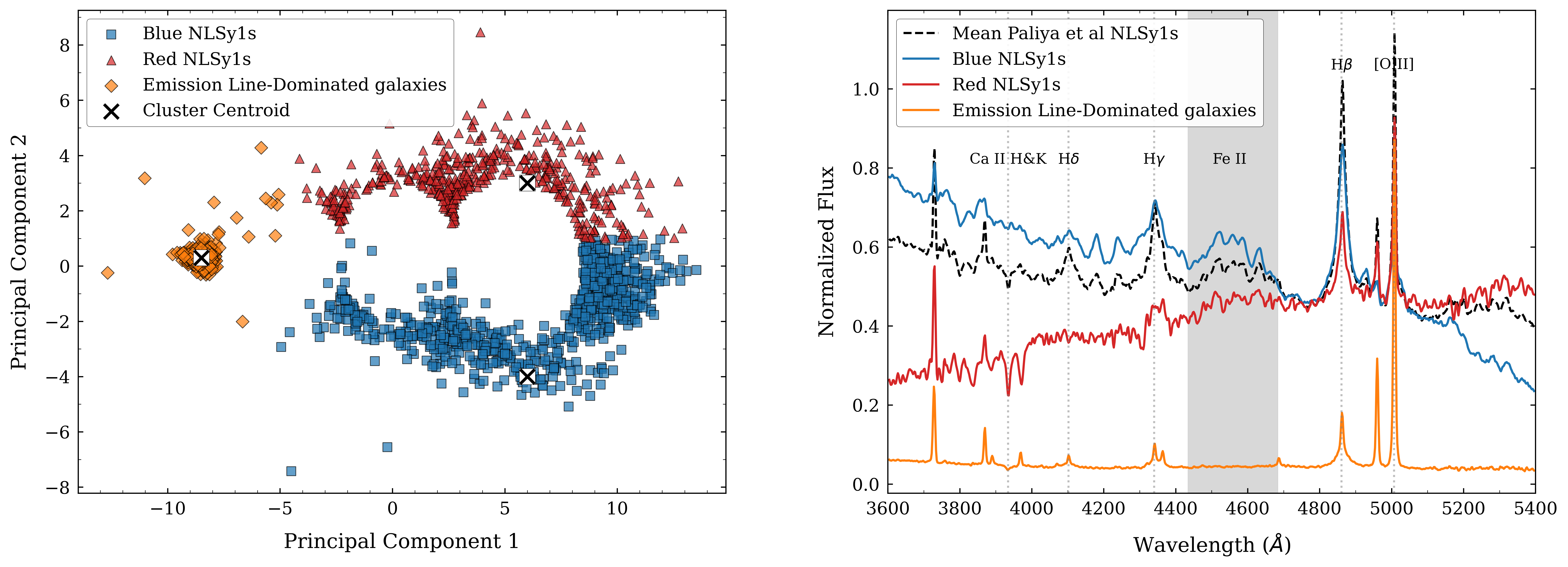}
    \caption{Left: A 2D projection (principal component (PC1) versus PC2 coefficients) of the anomalous galaxies as divided into three groups (group 1: orange, group 2: blue, group 3: red), in the 20 dimensional PCA hyperspace, by the second k-means clustering applied only on the anomalous galaxies. The groups are characterized by further analysis (see Sect.\ref{sec: Physical Diagnostics}). Right: Mean composite spectrum of each anomaly group. The colors of the spectra correspond to the color of their group in the left panel. The black dashed spectrum is the mean composite of all the galaxies in the \cite{paliya2024narrow} catalog. The red, blue and black spectra are arbitrarily normalized for better comparison.}
    \label{fig:Anomalies and Cluster}
\end{figure*}

%###################################################################################
%###################################################################################

\section{Results} \label{Results}
Hereafter, the three classifications done by the initial k-means clustering applied to the complete dataset will be referred to as clusters, while the classification of the detected anomalous spectra from the subsequent k-means clustering will be called groups.

\subsection{Clusters}

The initial k-means clustering was performed on the PCA-transformed dataset, and the resulting clusters were visualized using a two-dimensional projection of the PCA hyperspace. Specifically, we plotted the coefficients of the first two principal components, which capture the maximum variance in the data. The left panel of Fig.~\ref{fig:Cluster and Cluster Mean Spectra} displays the full quasar sample, segmented into three distinct clusters by the k-means algorithm. The identified spectral anomalies (see Sect.\ref{subsection:Anomaly Detection}) are overlaid as yellow star markers in the same figure for visual comparison.

To investigate the distinguishing features among the clusters, we computed and normalized the mean spectrum for each cluster (see right panel of Fig.~\ref{fig:Cluster and Cluster Mean Spectra}). The resulting mean spectra reveal that the primary differentiating factor is the strength of the emission lines, rather than the line widths or continuum shape. Cluster 1 exhibits the most prominent emission lines (e.g. average \oiii equivalent width (EW) $\approx36.26$, \hb EW $\approx91.15$), while cluster 3 shows the weakest (e.g. average \oiii EW $\approx2.35$, \hb EW $\approx71.92$), with cluster 2 falling in between (e.g. average \oiii EW $\approx10.55$, \hb EW $\approx79.28$). This trend is particularly pronounced in the \oiii and \oii emission lines. In cluster 1, the \oiii line is notably narrow and prominent, indicative of strong and possibly low-velocity narrow-line region emission. A similar but less pronounced pattern is observed in the Balmer lines. Interestingly, the trend is reversed for \feii emission, with cluster 3 displaying the strongest \feii emission features and cluster 1 the weakest. This inverse relationship between \feii strength and high-ionization narrow-line emission (e.g., \oiii) is consistent with known AGN spectral sequences (e.g., Eigenvector I of \cite{1992ApJS...80..109B}), and may reflect differences in accretion rate, ionizing continuum shape, or orientation effects across the clusters \citep[e.g.][]{2014Natur.513..210S,marziani2001searching}. 

The number of members and anomalies in each cluster is given in Table \ref{table:Anomalies per cluster}:
\begin{table}[h!]
\caption{Number of members and detected anomalies in each cluster.}
\label{table:Anomalies per cluster}
\centering
\begin{tabular}{|c | c | c | c|} 
 \hline
 Datapoint & Cluster 1 & Cluster 2 & Cluster 3 \\ [0.5ex] 
   \hline
 Members & 7244 & 6518 & 8746 \\ 
 Anomalies & 312 & 384 & 298 \\
 \hline
\end{tabular}
\end{table}

%#######################################################################
\subsection{Groups}

As outlined in Sect.\ref{subsection: SQuAD Algorithm}, we performed a second k-means clustering exclusively on the subset of anomalous galaxies. This clustering yielded three distinct groups, visualized in the left panel of Fig.~\ref{fig:Anomalies and Cluster} as a two-dimensional scatter plot based on the coefficients of the first two principal components of PCA. The right panel of the same figure displays the mean composite spectra corresponding to each of the three anomaly groups, with colors matching their cluster assignments. In contrast to the relatively similar continuum of the mean spectra of the clusters (see Fig.~\ref{fig:Cluster and Cluster Mean Spectra}), the composite spectra of the anomaly groups exhibit substantial diversity in both continuum slope and emission line characteristics. 

\subsubsection{Group 1}
The first group is characterized by galaxies exhibiting exceptionally strong emission lines superimposed on an almost flat continuum. These objects occupy the leftmost region in the PCA component space, shown as orange scatter points in the left panel of Fig.~\ref{fig:Anomalies and Cluster}. Their spectra are visually striking—dominated entirely by sharp and strong atomic emission lines with little to no underlying continuum emission (see orange spectra in the right panel of Fig.~\ref{fig:Anomalies and Cluster}). These spectral features are characteristic of Intermediate Seyfert galaxies, a detailed analysis of which is presented in Sect.~\ref{sec: Physical Diagnostics}. The absence of a prominent continuum enhances the visibility of numerous weak atomic lines that are typically obscured in conventional AGN and NLSy1 spectra, making this group particularly unique. The spectral shape and continuum-emission line contrast features are remarkably similar to the \civ peakers identified by \cite{tiwarisquad1}. 

Notably, this group is mildly affected by instrumental artifacts, most commonly in the form of sharp, narrow spikes that resemble strong, unresolved emission lines. These features—likely due to cosmic ray hits or processing glitches—often mimic the narrow \oiii profile that is characteristic of this group, leading to coincident grouping. However, these contaminant spectra were few in number and were manually identified and removed during visual inspection. In total, 257 such galaxies were identified, constituting approximately 29.3\% of the full set of detected outliers.

%#######################################################################
\subsubsection{Group 2}
The second group comprises galaxies characterized by a noticeably bluer optical continuum compared to the average (mean composite) NLSy1 spectrum, along with a significantly enhanced optical \feii emission bump. These objects populate the lowermost region of the PCA component space, as indicated by the blue scatter points in the left panel of Fig.~\ref{fig:Anomalies and Cluster}. The second principal component (PC2) coefficients primarily characterize spectral reddening effects, as demonstrated by \cite{tiwarisquad1}. Hence, their presence at the low end of the PC2 axis reflects a lower degree of continuum reddening or intrinsically bluer spectral slopes. Owing to these characteristics, we refer to this population as blue NLSy1s.

Spectroscopically, the \blues exhibit the strongest optical \feii emission, with an average R4570 $\approx0.647$ as compared to $\sim0.532$ for \reds and $\sim0.103$ for \ldnls. The \blues also have the weakest \oiii $\lambda5007$ emission among the three anomaly groups (see blue spectrum in the right panel of Fig.~\ref{fig:Anomalies and Cluster}) with average \oiii EW $\approx5.12$ \AA\,as compared to $\sim 8.28$\AA\,for reds and $\sim165.35$\AA\,for \ldnls. This inverse correlation between \feii and \oiii strength is a hallmark of the so-called eigenvector 1 parameter space, which captures one of the principal axes of quasar spectral diversity and is commonly linked to variations in Eddington ratio, orientation, and broad line region (BLR) structure \citep{1992ApJS...80..109B}. In total, 374 such anomalies were identified, accounting for 42.64\% of the detected anomalies and constituting the largest group in the anomaly sample.
%#######################################################################
\subsubsection{Group 3}

The third group consists of NLSy1 galaxies that exhibit a markedly redder optical continuum compared to the average NLSy1 spectrum. These objects are concentrated in the upper region of the PCA projection (see red scatter points in the left panel of Fig.~\ref{fig:Anomalies and Cluster}), corresponding to elevated PC2 values. High values along this principal component reflect a steep positive continuum slope, consistent with significant reddening. We refer to this class hereafter as red NLSy1s.

The mean spectrum of the \reds reveals moderate \oiii emission and relatively weak \feii features (see red spectrum in the right panel of Fig.~\ref{fig:Anomalies and Cluster}), positioning them between the \blues and the \ldnls in terms of narrow-line strength and iron emission. However, their most distinctive attribute is the reddened continuum, which gives rise to their classification. 
In total, we identified 246 such red NLSy1s, accounting for 28.05\% of the detected anomalies.

The number of anomalies in each group is given in Table \ref{table:Anomalies per group}.

\begin{table}[h!]
\caption{Number of anomalies in each group after manual removal of bad spectra caused due to instrumental artifacts.}
\label{table:Anomalies per group}
\centering
\begin{tabular}{|c | c | c | c|} 
 \hline
  Group & Group 1 & Group 2 & Group 3 \\ [0.5ex] 
   \hline
 {\tt\string Count} & 257 & 374 & 246 \\ 
 \hline
\end{tabular}
\end{table}

%#######################################################################
%#######################################################################
\section{Discussion} \label{sec: Physical Diagnostics}
The \citet{paliya2024narrow} catalog was curated by selecting sources solely on the basis of quantitative definition markers of an NLSy1 galaxy. This makes it possible for contaminants to seep into the sample that are not actually NLSy1 galaxies but fall into the numerical definition. Since we are dealing with outliers from the \citet{paliya2024narrow} catalog, it becomes specially important to confirm the true classification of these sources before proceeding with an in-depth spectral and physical analysis. By careful spectral decomposition of the detected outliers using PyQSOFit \citep{2018ascl.soft09008G} we found that two of the three anomaly groups i.e. \reds and \blues are truly NLSy1s. We define a true NLSy1 to be a Seyfert 1 galaxy with \hb FWHM less than 2000 \kms, \oiii to \hb ratio $<3$, strong \feii emission and (in our case) a broad \hb emission profile with negligible to no narrow component. Group 1 members, on the other hand were found to be Intermediate Seyfert galaxies. Their nature was confirmed by (i) the presence of a composite (narrow+broad) profile for H$\beta$ and H$\alpha$ emission lines \citep{cohen1983narrow}, (ii) \hb narrow to \hb broad flux ratio of $\approx0.2$ as compared to average ratio $\sim0.04$ for the total sample. We also followed the classification scheme provided by \citet{whittle1992virial} and found that out of 247 sources, 47 fall in the Seyfert 1.2 category while the rest can be classified as Seyfert 1.5. The Seyfert type calculation as given by \citet{netzer1990agn} also provided a similar distribution, confirming the Intermediate Seyfert identity of Group 1 members. For sources with low signal to noise ratio (SNR),for which we were unable to obtain a reliable classification were left unmarked but are still given in the catalog for future observations.

To gain a deeper insight into the nature of the three groups—we carried out a comprehensive investigation of their physical properties. This includes an examination of their redshift distributions to probe potential selection effects and cosmological trends, estimation of color excess $E(B-V)$ as a proxy for internal reddening, infrared (IR) color diagnostics to assess dust attenuation and thermal emission properties, and orientation-dependent effects inferred through the EW of the \oiii$\lambda5007$ line. All the parameters used in the following analysis, such as M\textsubscript{BH}, bolometric luminosity, line properties etc. are taken from \citet{paliya2024narrow} catalog. These diagnostics collectively allow us to characterize the underlying physical drivers responsible for the observed spectroscopic diversity among the anomaly classes and to place them in the broader context of AGN unification and evolution.

\begin{figure}[h!]
    \centering
    \includegraphics[width=1\linewidth]{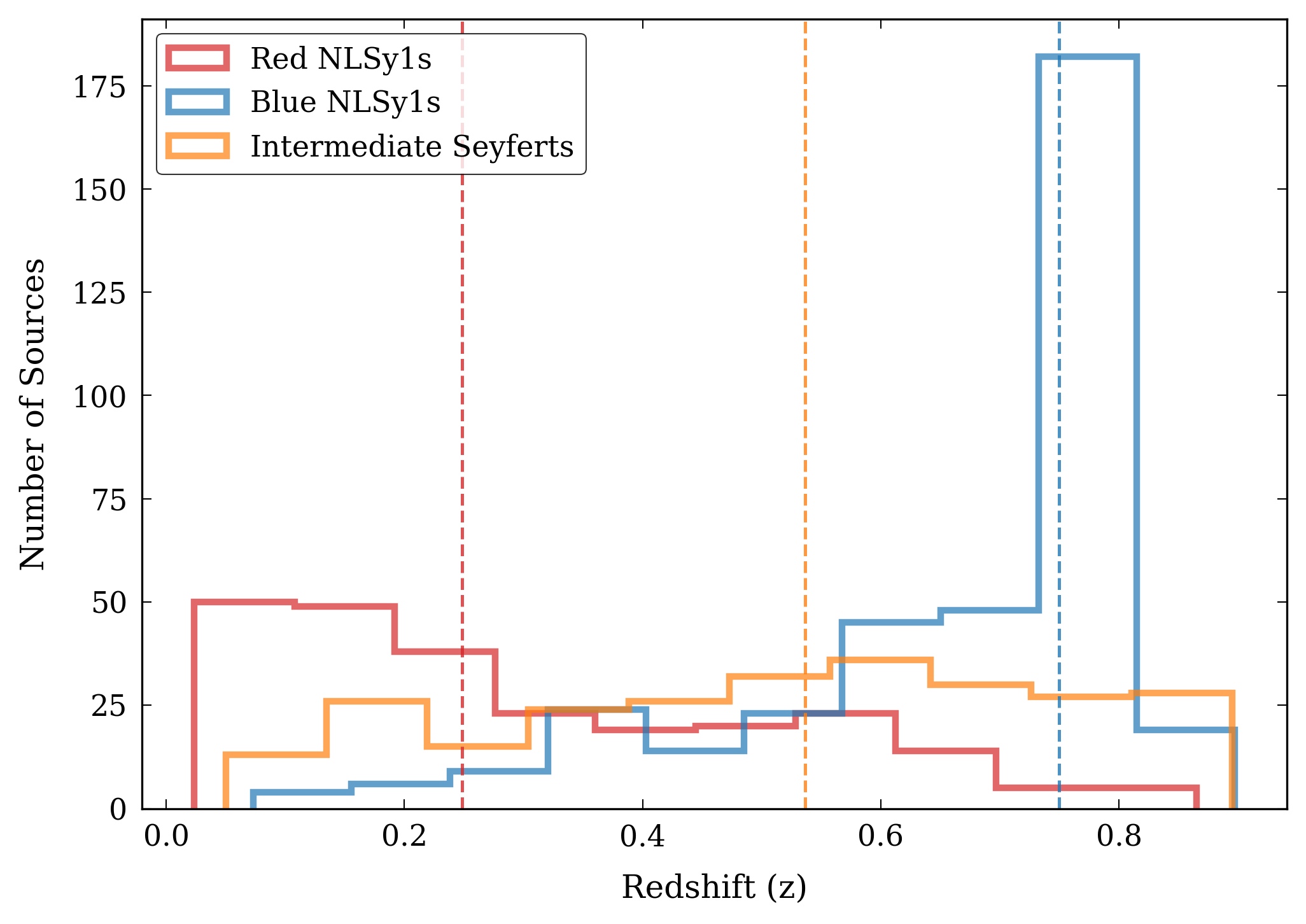}
    \caption{The redshift distribution of red (red), blue (blue) NLSy1s and Intermediate(orange) Seyferts. The median redshift for each class is marked by a dashed line.}
    \label{fig:redshift distribution}
\end{figure}

%#######################################################################

\subsection{Redshift and magnitude distribution} \label{sec: redshift and magnitude}

Fig.~\ref{fig:redshift distribution} illustrates the redshift distribution of the three identified anomaly types. The \reds predominantly occupy the lowest redshift range, with their numbers declining steadily as redshift increases. In contrast, the \blues are more prevalent (infact heavily concentrated at $z\sim0.8)$ at higher redshifts and are virtually absent at $z\sim0$. The \ldnls span an intermediate redshift range and exhibit a relatively uniform distribution across the redshift space.

\begin{figure}[h!]
    \centering
    \includegraphics[width=1\linewidth]{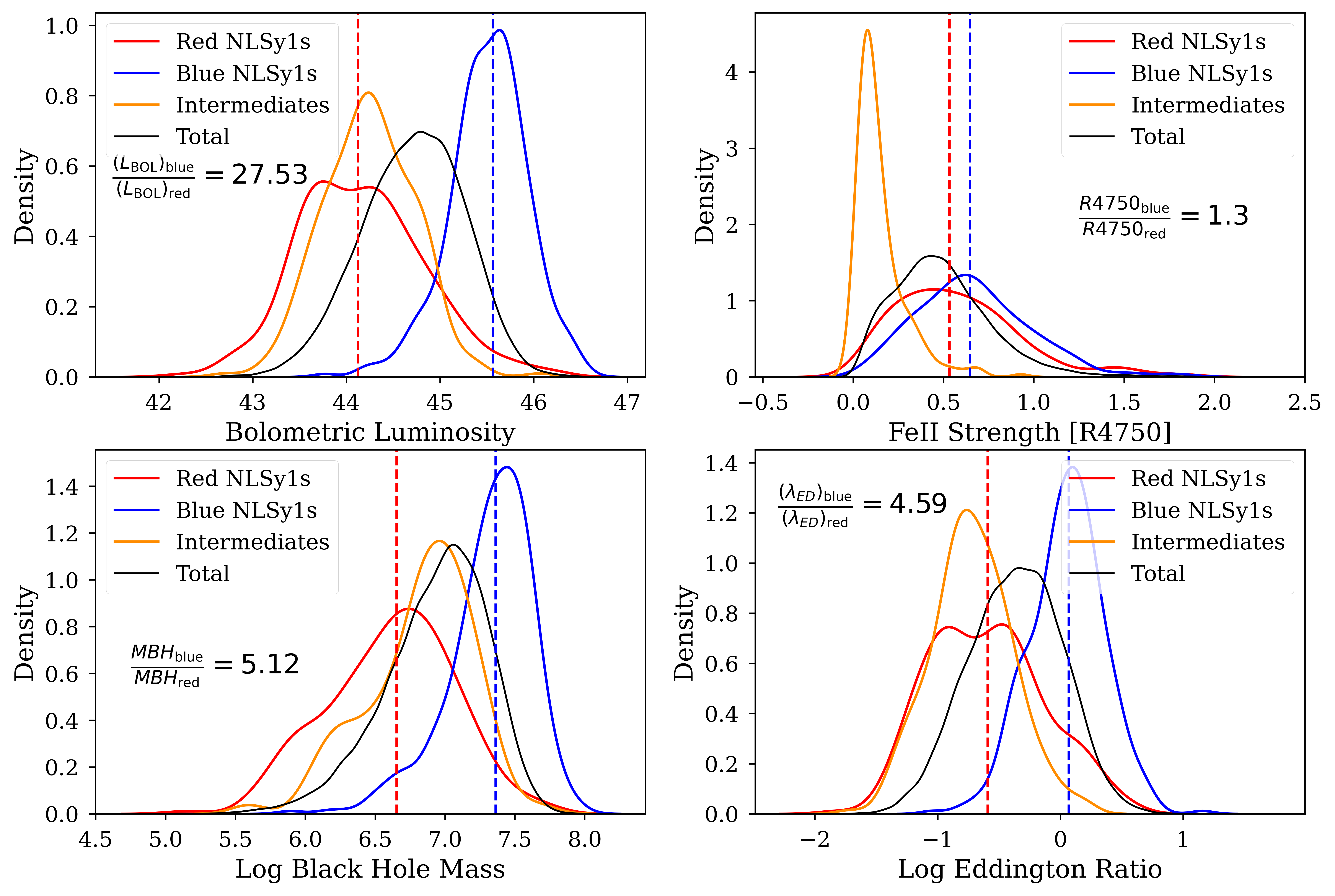}
    \caption{The density normalized distribution of parameters for the total \protect\cite{paliya2024narrow} dataset and the three anomalous groups. Top Left: The distribution of Eddington ratio on log scale. Top Right: Distribution of \hb derived black hole mass on log scale. Bottom Left: log scale distribution of bolometric luminosity and; Bottom Right: Distribution of \feii emission strength (R4750). In all the cases, the \reds lie at the lowest end and \blues at the highest. \ldnls are the galaxies with the lowest \feii emission strength in the complete dataset. The number annotation in each plot indicates the ratio of the respective value for blue to red NLSy1s.}
    \label{fig: derived properties}
\end{figure}

The decline in the number of \reds with increasing redshift is likely not intrinsic to their astrophysical properties but rather a consequence of observational bias, primarily driven by the sensitivity limitations of the SDSS eBOSS spectrograph \citep{lupton2002sdss}. The \reds are observed to exhibit low luminosity (see lower left panel of Fig.~\ref{fig: derived properties}), and as the redshift increases, their apparent brightness diminishes significantly due to cosmological dimming \citep[see][]{calvi2014effect} and increasing luminosity distance. Consequently, beyond a certain redshift threshold, many \reds fall below the SDSS detection limit and are thus underrepresented in the sample. This selection effect creates the appearance of a strong redshift dependence, when in fact simply reflects the survey’s inability to detect faint sources at greater distances. Therefore, any inferred evolutionary trends or redshift-dependent population changes among the \reds must be interpreted with caution, as they may be dominated by this selection bias rather than by underlying astrophysical evolution.

On the other hand, the scarcity of \blues at low redshifts can be understood in the context of the evolution of the quasar luminosity function \citep[QLF;][]{shen2020bolometric,hopkins2007observational}. Quasar activity is known to peak at redshifts around $z\sim2$, often referred to as the "quasar epoch," and declines both toward lower and higher redshifts (e.g. \citealt{boyle20002df, richards2006sloan}). As a result, the probability of detecting luminous quasar-like objects such as the \blues is significantly higher at intermediate to high redshifts. This naturally leads to a larger observed population of \blues at $z \gtrsim 0.7-0.8$, with their numbers falling off as redshift decreases. The drop-off toward lower redshifts is therefore consistent with the global decline in quasar activity and abundance, as traced by the QLF.

Visual inspection of the SDSS images for the \reds and \blues revealed that majority of the \reds have a visible host galaxy. This is likely facilitated by two factors i) The \reds have low AGN luminosity making the center of the image not as bright as compared to the host hence making it possible to see the host and ii) The \reds are present primarily at low redshifts due to which the host galaxy is easily identifiable in the images. Notably, most of these host galaxies have a spiral morphology and exhibited a bar \citep{2003AJ....126.1690C,2023A&A...679A..32V}. In contrast, \blues have a \say{blue dot} appearance, consistent with their higher AGN luminosities and larger redshifts, which render host galaxy detection challenging.

Interestingly, the redshift distribution of the \ldnls is nearly uniform throughout the observed redshift range with a median $z \sim 0.6$. Their relative absence at lower redshifts (as compared to \reds) can be reasonably attributed to the evolution of the QLF, which predicts a decline in AGN activity as one moves away from the peak epoch near $z \sim 2$ \citep{richards2006sloan}. However, the missing Intermediate Seyfert population at higher redshifts ($z > 0.6$; like \blues) appears to arise from selection effects inherent to the SDSS target selection algorithm. Specifically, the SDSS AGN selection is partly based on the observed $g - i$ color, where a minimum threshold is required for a source to be classified as a quasar candidate \citep{myers2015sdss}. As redshift increases beyond $z \sim 0.6$, the \hb and \oiii $\lambda\lambda4959,5007$ emission line complex—one of the most prominent features in the \ldnls spectra—shifts into the $i$-band filter. This leads to an artificial brightening of the $i$-band magnitude, which in turn suppresses the $g - i$ color. As a result, some of these objects fall below the quasar classification threshold and are excluded from the SDSS spectroscopic quasar sample. This selection bias offers a compelling explanation for the subtle decline in the \ldnls population at higher redshifts, despite the intrinsic strength of their emission lines.

The top-left panel of Fig.\ref{fig: derived properties} shows the bolometric luminosity distribution of all the classes. The \blues exhibit the highest values lying mostly at the high end tail of the overall distribution. In contrast, the \reds and \ldnls occupy the lower region of the overall distribution and have a median bolometric luminosity lesser as compared to the entire sample. This bias in the luminosity of \blues is also due to the fact that only high luminosity sources would be visible at higher redshifts leading to this distribution. Additionally, due to this difference in the luminosity distribution, the connected properties i.e. the M\textsubscript{BH} and Eddington ratio in the lower panels of Fig.\ref{fig: derived properties} also show a similar trend, where \blues occupy the highest end. The \feii emission strength in the upper right panel of Fig.\ref{fig: derived properties} shows that the \reds and \blues have nearly similar \feii emission strengths with the \blues having a subtle excess. The \ldnls on the other hand have exceptionally low \feii emission, occupying the lowermost region of the overall distribution. This is consistent with \citet{cohen1983narrow}, as the work used weak \feii emission as a criterion for selecting Intermediate Seyfert samples.

\begin{figure}[h!]
    \centering
    \includegraphics[width=1\linewidth]{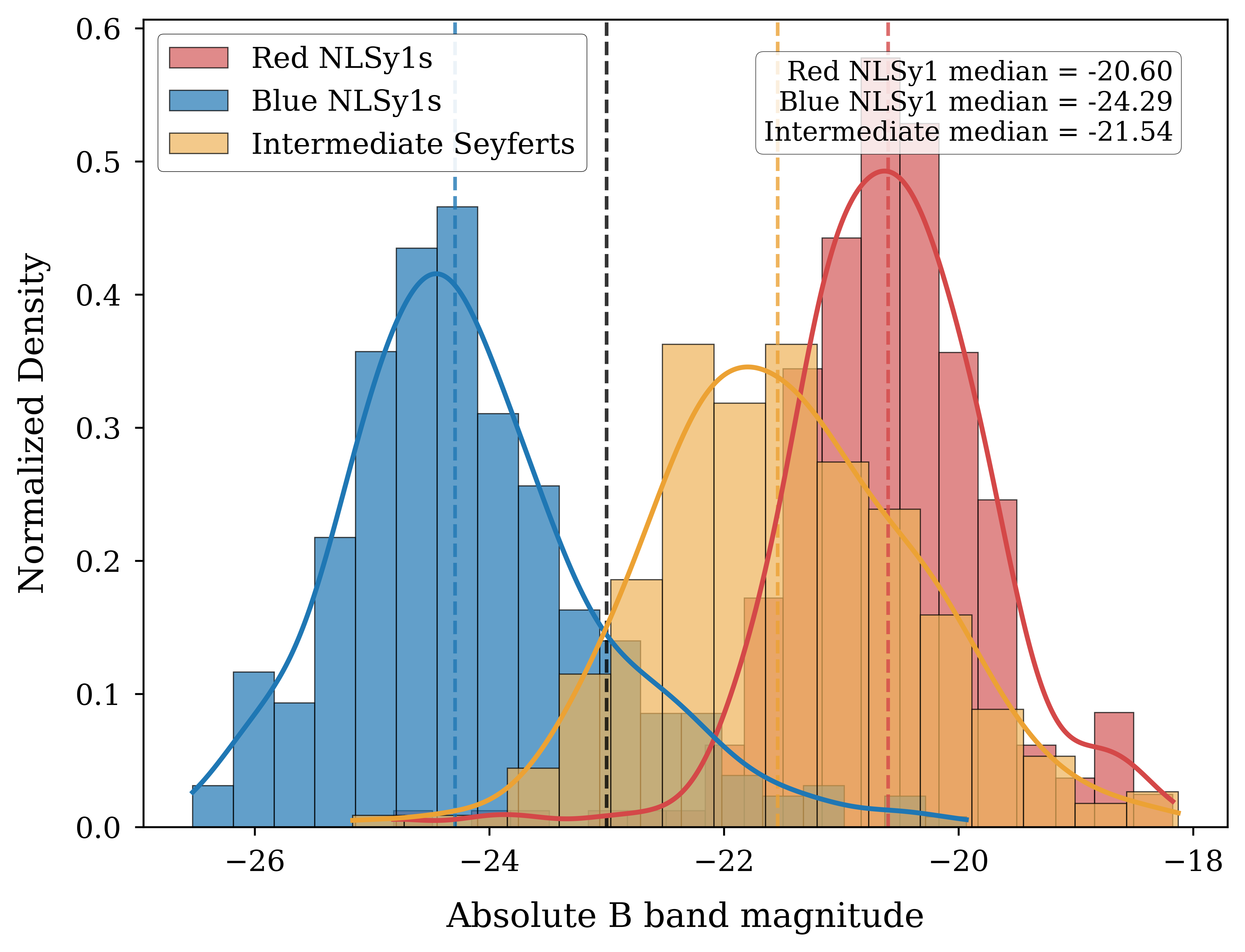}
    \caption{Absolute B band magnitude distribution for the red, blue NLSy1s and Intermediate Seyferts. The respective median is marked by dashed lines for both classes. The black dashed line marks a $M_B=-23$ which is considered to be a distinction limit between quasars and Seyfert galaxies. \protect\citep{paliya2024narrow}}
    \label{fig:MagB_Distribution}
\end{figure}

Additionally, as shown in Fig.~\ref{fig:MagB_Distribution}, most of the \blues exhibit very high values for the absolute B band magnitude (more than 23) as compared to the other two classes. This indicates that most of the blue NLSy1 sources are more like a quasar and less of a Seyfert galaxy pertaining to their strong luminosity. Red NLSy1s on the other hand, have the lowest B band magnitude which is consistent with their low AGN activity and host galaxy dominated emission. The \ldnls also lie towards the lower end of the distribution with a median B band magnitude slightly higher than that of red NLSy1s.\\

Given the reddened continuum and lower luminosity of the red NLSy1s, it is tempting to interpret them as dust-obscured AGN, potentially viewed through the edge of the torus \citep[for eg.][]{li2015detection,zhang2017reddening,pan2017candidate}. This interpretation was debunked as the reddening was found to be caused by host galaxy dominance instead of obscuration (see Sect.\ref{subsec:ebv}). On the other hand, the \blues—with their bright, blue continua—appear to be face-on sources with minimal extinction or attenuation. Such an interpretation would imply that the observed differences in bolometric luminosity between these classes could be driven primarily by orientation effects or dust obscuration, rather than intrinsic differences in AGN power. To critically assess this possibility, we conduct a series of diagnostic analyses aimed at disentangling the contributions of orientation, dust attenuation, and intrinsic AGN luminosity. These studies serve to clarify whether the observed spectral and luminosity differences among the three anomaly types—red NLSy1s, blue NLSy1s, and \ldnls—are the result of physical distinctions or observational biases.

%#######################################################################
\subsection{Color excess: E(B-V)}\label{subsec:ebv}

Dust obscuration in AGNs can be quantified using the color excess, E(B-V), which measures the reddening of the continuum due to dust absorption \citep{calzetti2000dust,maiolino2001dust}. In order to calculate the E(B-V) values, all spectra were de-reddened to account for the Milky Way dust extinction \citep[][]{1999PASP..111...63F} using the dust maps by \cite{Schlegel_1998}. Then we fit this de-reddened spectra $f(\lambda)$ with a template spectra (\citet{2001AJ....122..549V} composite) $f_t(\lambda)$ using the parameterization following \cite{2012MNRAS.423.2879V}:
\begin{equation} \label{equation: ebv parameterization}
    f(\lambda) = \left[af_t(\lambda) + b\left(\frac{\lambda}{\lambda_0}\right)^\alpha\right]e^{-\tau_\lambda}
\end{equation}

\begin{figure}[h!]
    \centering
    \includegraphics[width=1\linewidth]{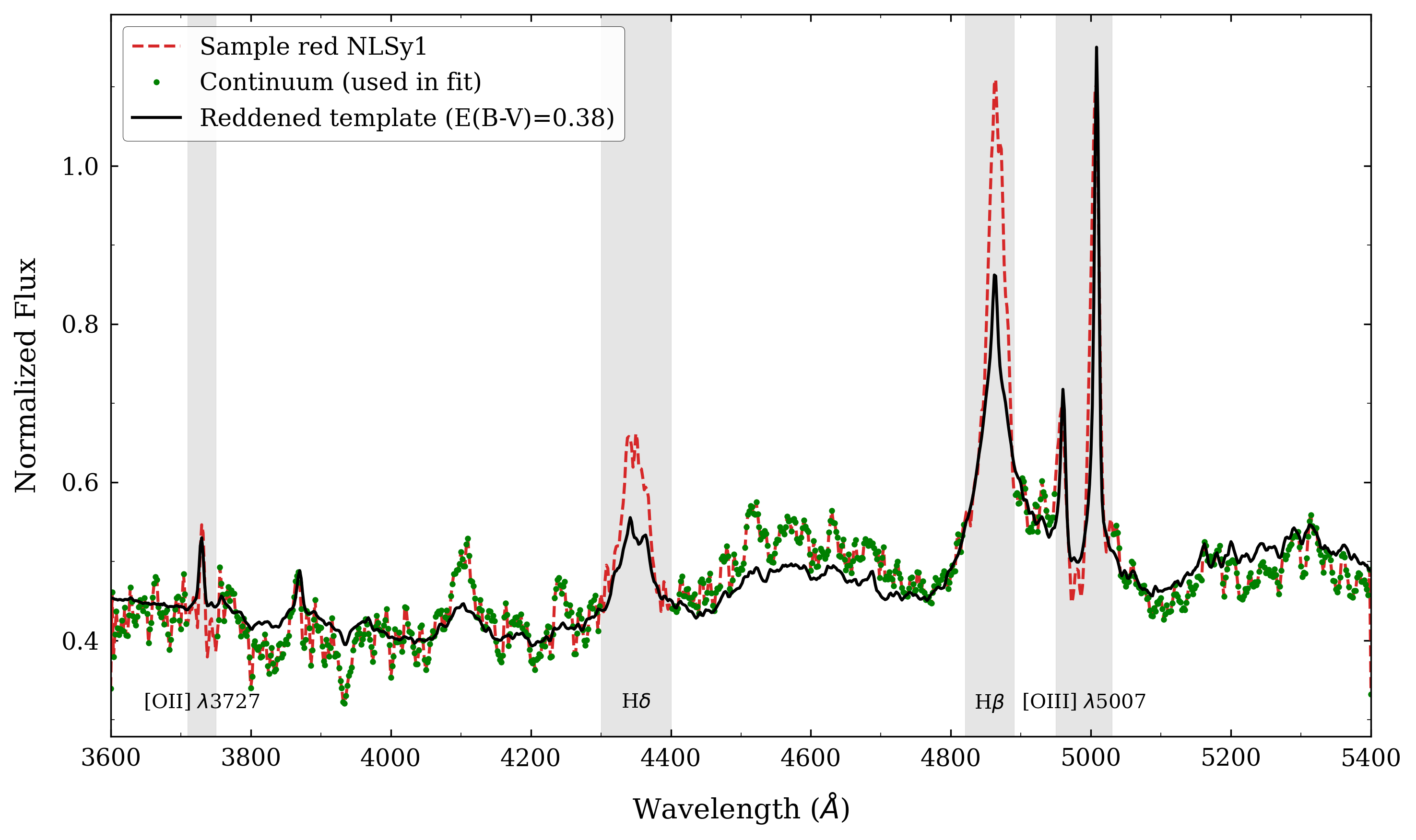}
    \caption{Sample E(B-V) calculation for an NLSy1 spectrum using $\protect\chi^2$ minimization. The red line shows the reddened \protect\cite{2001AJ....122..549V} composite spectrum matched to the Milky Way extinction corrected NLSy1 spectrum (blue). The gray shaded regions mark the emission line regions masked during $\chi^2$ calculation.}
    \label{fig: EBV Sample}
\end{figure}

We utilize a $\chi^2$ minimization (masking out the emission line regions, marked as gray in Fig.~\ref{fig: EBV Sample}) to obtain the optimum values for parameters $a$, $b$, $\alpha$, $\lambda_0$ and $\tau_\lambda$. The first term in Eq.\ref{equation: ebv parameterization} gives the scaling difference, while the second term denotes the difference in spectral index between the observed and template spectra. The optical depth for dust \say{$\tau_\lambda$} is obtained for the Small Magellanic Cloud (SMC) like extinction curve \citep[][]{gordon2003quantitative}. The above described analysis was done using {\tt\string dust\_extinction} Python package \citep[][]{Gordon2024}.
\begin{figure}[h!]
    \centering
    \includegraphics[width=1\linewidth]{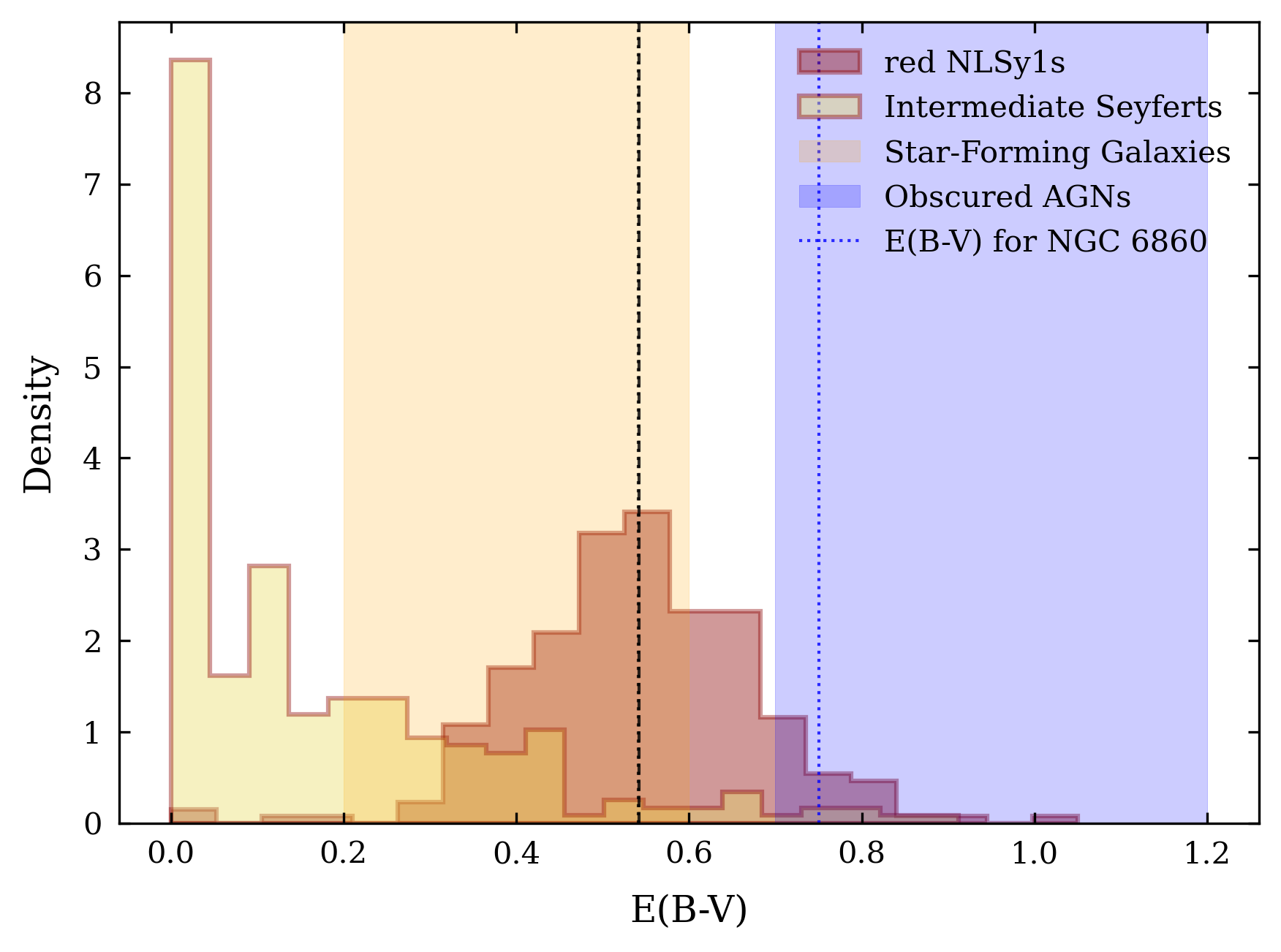}
    \caption{$E(B-V)$ distribution for the red NLSy1s and Intermediate Seyfert galaxies. The $E(B-V)$ values for all the \blues were calculated to be zero consistently and hence are not shown in the figure for visual clarity. The yellow shaded region indicates the typical \ebv range for star-forming galaxies while the blue region indicates dust-obscured AGNs. The black-dashed line indicates the median \ebv for the \reds at 0.54, falling inside the star-forming range. The blue-dotted line indicates the \ebv=0.72 for NGC 6860 \protect\citep[a known dust-obscured AGN;][]{lipari1993high}.}
    \label{fig:EBV_distribution}
\end{figure}

As illustrated in Fig.~\ref{fig:EBV_distribution}, the distribution of $E(B-V)$ values for the \ldnls is majorly concentrated around zero and spans up to 0.5. On the other hands, the distribution for \reds primarily spans from 0.3 to 0.8, with a median of 0.54. The $E(B-V)$ values for all the \blues were calculated to be zero consistently and hence are not shown in Fig.~\ref{fig:EBV_distribution} for visual clarity. Several previous studies provide context for interpreting these $E(B-V)$ values in terms of host galaxy effects, reddening and AGN activity. For example, analyses of SDSS galaxies in terms of extinction and star formation histories by \citet{kauffmann2003stellar} and \citet{brinchmann2004physical}, as well as investigations of dust attenuation in X-ray-selected NLSy1s by \citet{caccianiga2004xmm}, show that the $E(B-V)$ range for \reds is well within that of typical star-forming galaxies. On the other hand, the values for Intermediate Seyferts and \blues suggest little to no dust obscuration. Broader studies of dust extinction in galaxies across various redshifts further support this interpretation \citep{garn2010predicting, buat2011goods}.

Crucially, the moderate $E(B-V)$ values observed in the \reds are insufficient to classify them as dust-obscured AGN. For instance, the well-studied obscured AGN, NGC 6860 exhibits a significantly higher $E(B-V) \sim 0.72$, as reported by \citet{lipari1993high}. More recent work by \citet{andonie2025obscured}, based on SED fitting of obscured AGN, and theoretical extinction models from \citet{schnorr2016feeding} suggest that $E(B-V)$ values in the range of 0.7–1 are indicative of significant dust obscuration, while values above 1 are typical of fully extinguished AGN. The fact that the \reds fall just below this threshold supports the interpretation that their reddened appearance is not primarily due to heavy dust attenuation. Hence, the redder appearance is more plausibly attributed to intrinsic properties of the host galaxy, such as a weak AGN continuum being overwhelmed by starlight or modest dust distributed on galactic rather than nuclear scales.

\begin{figure}[h!]
    \centering
    \includegraphics[width=1\linewidth]{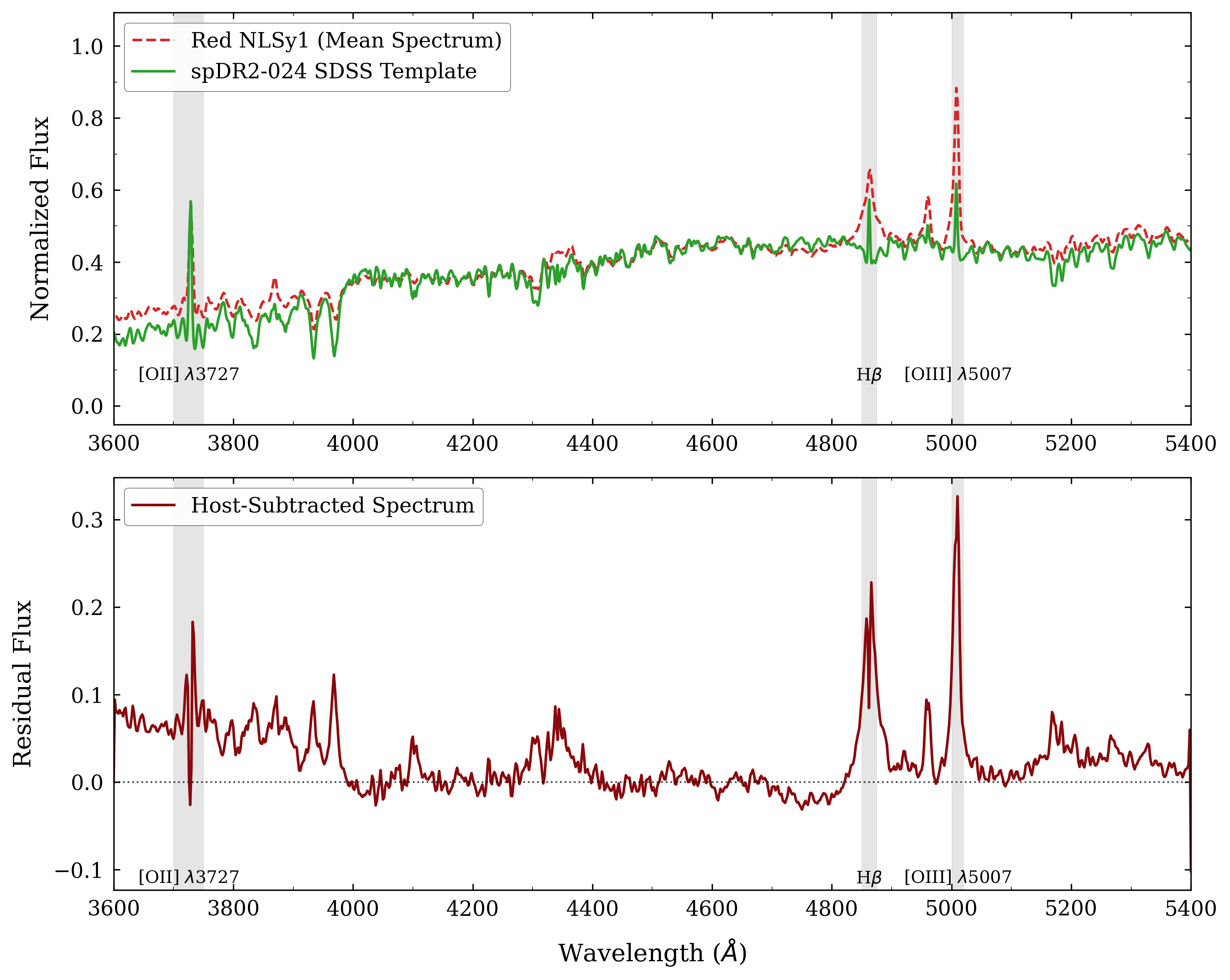}
    \caption{Top: The comparison of the mean \protect\reds spectrum (red) with a galaxy spectra template (green) from the SDSS. The template is arbitrarily normalized to best match the mean \reds spectrum. The AGN activity is apparent in the subtle continuum excess at the blue end and the broad \protect\hb and \protect\oiii emission lines. Bottom: Host galaxy subtracted spectrum for the \reds.}
    \label{fig:rednls1 versus galaxy template}
\end{figure}

As an additional measure to confirm the nature of the continuum reddening in the red NLSy1s, we compared the mean \reds spectrum with an SDSS galaxy template spectrum (see Fig.~\ref{fig:rednls1 versus galaxy template}). The comparison revealed a remarkably good match across the full wavelength range, indicating that the optical continuum of these sources is dominated by host galaxy starlight rather than AGN emission. The most notable deviations from the template arise in the broad emission lines—particularly H$\beta$ and \oiii—which are significantly wider in the \reds spectrum compared to those in the galaxy template. This confirms the presence of a BLR and narrow-line region (NLR) and validates the classification of these sources as Type 1 AGN despite the apparent host-dominated continuum. Such instances where the presence of AGN is only made evident by the emission lines have been studied in several literatures such as \cite{1996ApJ...457..199G}, \cite{kauffmann2003host} and the references therein.

A subtle excess in flux is also observed at the blue end of the spectrum, suggesting a weak but non-negligible AGN contribution that becomes more apparent at shorter wavelengths where AGN activity overpowers the galactic emission. In this scenario, the AGN is intrinsically weak in the optical band, and the host galaxy effectively overshines the nuclear continuum across most of the spectrum \citep[for eg.][]{ho1997search,ho2008nuclear,fernandez2023compact}. This interpretation is further supported by the clear presence of the 4000 \AA\,break (D4000) and the \caii H and K absorption doublet—prominent features characteristic of a galactic stellar population \citep[see][]{poggianti1997indicators}, and typically absent or diluted in AGN-dominated spectra. Additionally, the \ebv values for the host-galaxy subtracted spectra for the \reds were found to be zero consistently, indicating that nearly all the reddening was being contributed by the galactic emission.

Therefore, the overall agreement between the galaxy template and the observed continuum, along with these stellar absorption features and moderate $E(B-V)$ values, strongly suggest that the red colors of these NLSy1s are likely not driven solely by dust-induced extinction, but by a combination of weak AGN emission and strong stellar light from the host galaxy. This suggests that \reds occupy a regime where AGN activity is either intrinsically faint or temporarily diminished, allowing the galaxy’s stellar population to dominate the spectral energy distribution. Such systems may represent a low-luminosity end of the NLSy1 population or possibly an evolutionary stage where AGN activity is starting (young AGNs), but isn't strong enough to wash out the host galaxy spectral signature.

Additionally, it is interesting to note that for the \reds and blue NLSy1s, the H$\beta$ emission line was found to be better fitted with a Lorentzian profile (two Gaussian components used to fit the broad component while using PyQSOFit; while for the \ldnls a Gaussian profile was better suited. \citet{2020CoSka..50..270B} suggest that a Lorentzian profile is indicative of a more pronounced turbulent motion often associated with young sources. This is consistent with the \reds being galaxies where AGN activity has just began. On the other hand, the Gaussian profile for the \ldnls is supposed to indicate a class of more evolved objects.

%#######################################################################

\subsection{\oiii as inclination indicator}

\begin{figure}[h!]
    \centering
    \includegraphics[width=1\linewidth]{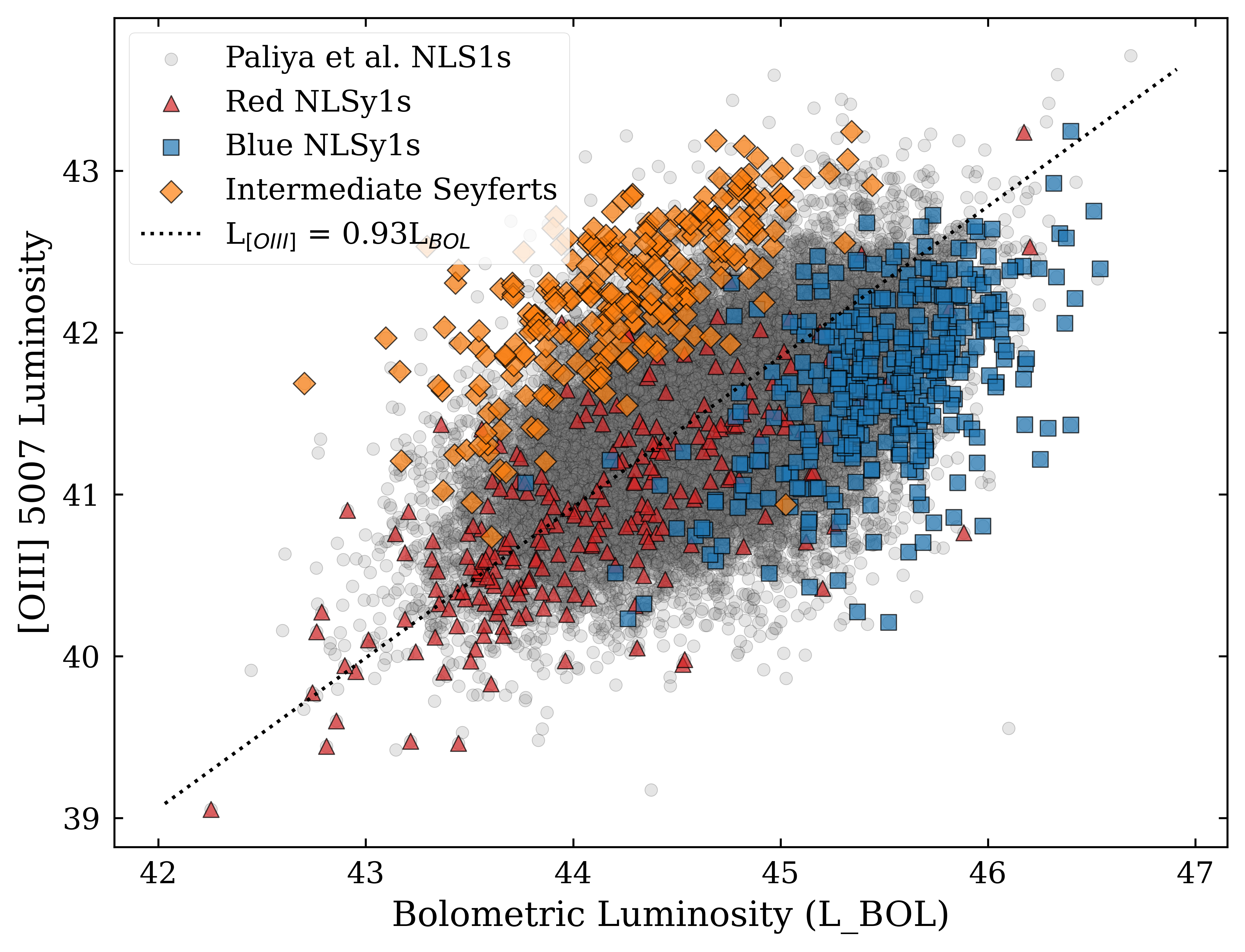}
    \caption{OIII Luminosity versus Bolometric Luminosity for the entire \protect\cite{paliya2024narrow} NLSy1 sample and the detected anomaly groups i.e \reds (red), \blues (blue) and \ldnls (orange). The black dotted line indicates 1:1 relation.}
    \label{fig:oiii-bolometric-luminosity}
\end{figure}

\begin{figure}[h!]
    \centering
    \includegraphics[width=1\linewidth]{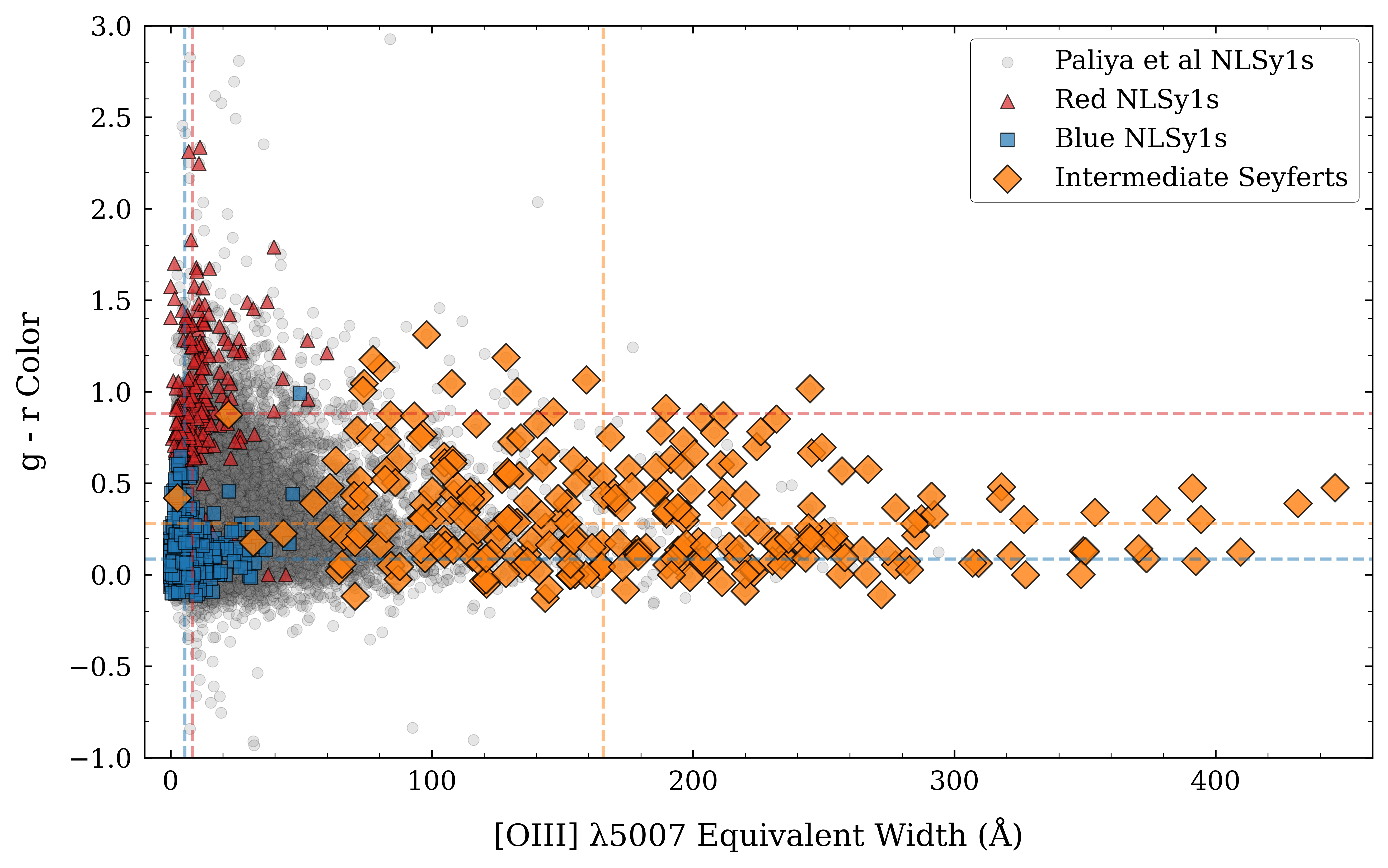}
    \caption{\oiii EW versus (g-r) optical color for the entire \protect\cite{paliya2024narrow} catalog and the three anomaly groups. The horizontal and vertical dotted lines indicate the median (g-r) and \oiii EW values, respectively, for each type.}
    \label{fig:g-r_vs_OIII}
\end{figure}

To evaluate the role of orientation in shaping the observed properties of our anomaly types, we reproduced the analysis of \citet{10.1111/j.1365-2966.2010.17843.x}, where they modeled the distribution of \oiii EW as a power law shaped by orientation-dependent projection effects. A prerequisite for this analysis is that \oiii luminosity serves as a reliable proxy for bolometric luminosity. As shown in Fig.~\ref{fig:oiii-bolometric-luminosity}, we find a strong correlation between these quantities (r-value = 0.731 and p-value = $1.89\times10^{-42}$), with a best-fit slope of 0.93, supporting this assumption. Interestingly, the \oiii luminosities of the \reds and \blues slightly underestimate the bolometric luminosity, while for the Intermediate Seyferts, the \oiii luminosity tends to overestimate it—suggesting intrinsic differences in their narrow-line region properties or covering factors.\\
The convolution framework of \citet{10.1111/j.1365-2966.2010.17843.x} involved fitting the log distribution of the \oiii equivalent width with a double gaussian from 2 to 30 \AA\,and a power-law from 30 to 500 \AA. We performed the same analysis for our NLSy1 sample, and found that it yields a best-fit power-law slope of $\Gamma = -2.55 \pm 0.1$. This was significantly shallower than the $\Gamma = -3.55$ reported for an orientation-unbiased quasar sample. This lower $\Gamma$ value likely indicates that our sample is restricted to lower inclination angles—either due to selection biases or intrinsic obscuration limits—and thus omits high-inclination sources where the torus obscures the central engine, naturally aligning with the Type 1 AGN classification. This supports the conclusion drawn from the color excess analysis in Sect.\ref{subsec:ebv}, which suggests a lack or negligible presence of dust obscuration. Moreover, the \oiii EW distributions of the \reds and \blues are statistically indistinguishable (see Fig.~\ref{fig:g-r_vs_OIII}), implying no significant inclination difference between them. The distinct optical colors ([u–g], [u–r], [g–r], [g-i]) between these two classes must therefore arise from intrinsic differences rather than geometric effects as discussed earlier. In contrast, the \ldnls show the highest \oiii EWs among the three types, suggesting that they are being observed at relatively higher inclinations consistent to their inherent nature.

\subsection{WISE-color analysis}
While several papers use Wide-field Infrared Survey Explorer (WISE) colors to distinguish AGN types or study dust obscuration \citep[for eg.][]{nikutta2014meaning,wu2012sdss}, the direct use of WISE color as a pure orientation indicator is debated. \cite{klindt2019fundamental} specifically argues that differences in WISE color distributions between red and blue quasars are not solely due to orientation, but also to evolutionary effects. \cite{fawcett2020fundamental} have used the WISE color differences as a signature of radio and intrinsic property differences in red and blue quasars. 
\begin{figure}[h!]
    \centering
    \includegraphics[width=1\linewidth]{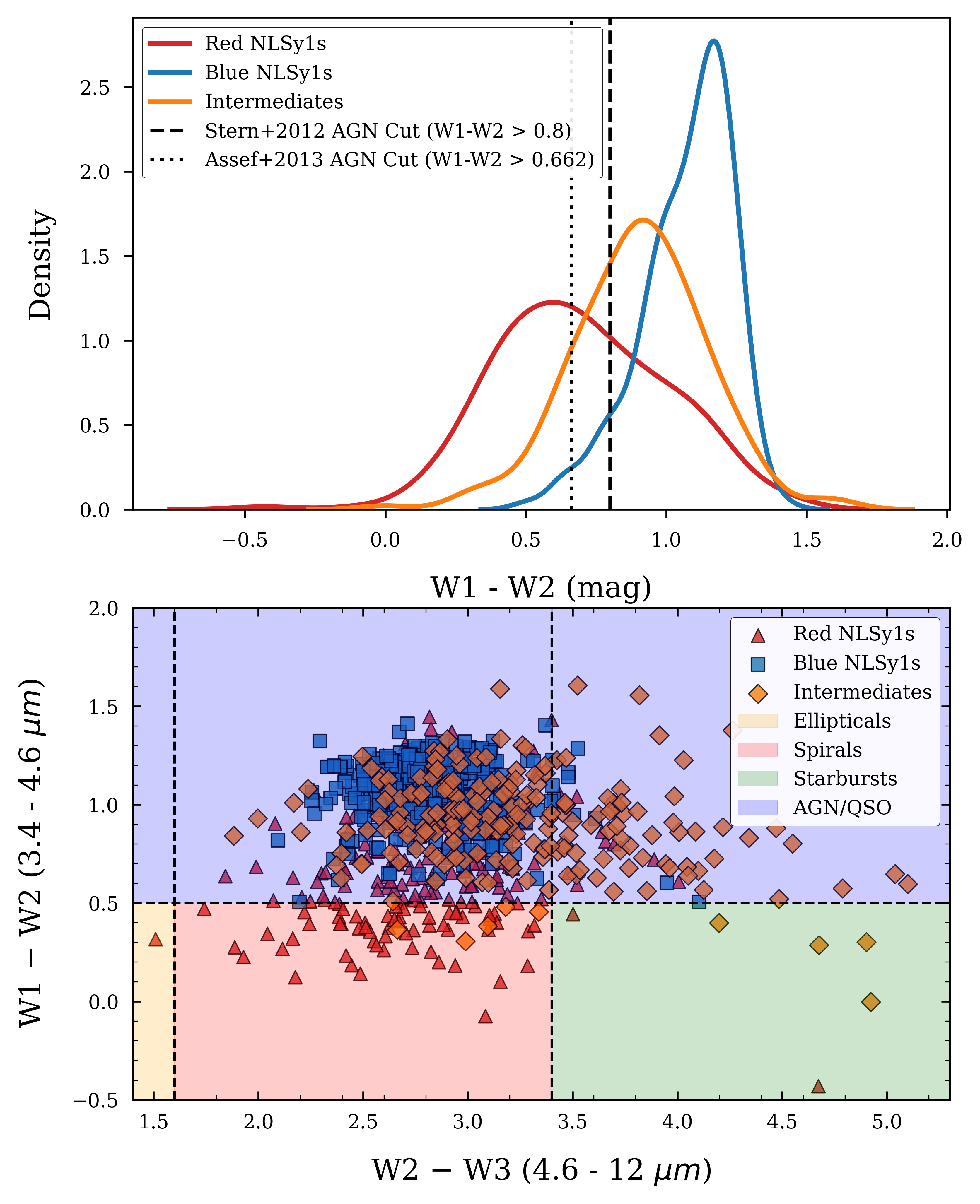}
    \caption{Top: Kernel density distribution of $W1 - W2$ colors for red NLSy1s, blue NLSy1s, and Intermediate Seyferts. Vertical lines indicate AGN selection thresholds from \cite{stern2012mid} and \cite{assef2013mid}. Bottom: WISE color-color diagram showing $W1 - W2$ versus $W2 - W3$ with shaded regions denoting different galaxy/AGN types (adapted from \citet{mingo2016mixr}).}
    \label{fig:Wise Analysis}
\end{figure}

Based on the WISE color analysis shown in Figure~\ref{fig:Wise Analysis}, we investigate the mid-infrared properties of three classes: red NLSy1s, blue NLSy1s, and Intermediate Seyferts. The top panel shows the kernel density distribution of $W1 - W2$ colors for these sub-classes. The blue NLSy1s exhibit a systematically higher $W1 - W2$ color peak (centered near $\sim$1.1 mag) than both \reds NLSy1s and Intermediate Seyferts, with the latter two showing broader distributions. This is particularly intriguing, as the $W1 - W2 \geq 0.8$ color cut (black dashed line) from \citet{stern2012mid} and $W1 - W2 \geq 0.662$ from \citet{assef2013mid} are commonly used AGN selection criteria. A majority of \blues clearly lie above both thresholds, implying robust AGN dominance in their mid-infrared emission. This is consistent with the \say{quasar} like properties of \blues, indicated by their absolute B band magnitude being greater than 23, as discussed in Sect.\ref{sec: redshift and magnitude}. Conversely, a significant fraction of \reds fall below these cuts, suggesting increased host galaxy contamination or intrinsically weaker AGN emission in the mid-IR, consistent with dilution effects. The bottom panel of Fig.~\ref{fig:Wise Analysis} (a color-color diagram and thresholds adapted from \cite{mingo2016mixr}) reinforces this interpretation: \blues and many \ldnls occupy the AGN/quasar region (purple-shaded), whereas \reds are more scattered into starburst and spiral galaxy regions. These patterns collectively suggest that \blues exhibit stronger nuclear activity with more prominent hot dust emission, while \reds may either be more host-dominated or have dust obscuration from the galaxy that affects their mid-IR colors differently. \ldnls span a wide color space, reflecting potential diversity in their dust geometry or evolutionary phase.

%#######################################################################

\subsection{BPT Diagram}

The Baldwin, Phillips \& Terlevich (BPT) diagram \citep{1981PASP...93....5B} is a widely used diagnostic tool in extragalactic astrophysics to classify the dominant ionization mechanism in galaxies based on strong optical emission-line ratios. Typically plotted using $[ \mathrm{O\,III} ] \lambda5007/\mathrm{H}\beta$ versus $[ \mathrm{N\,II} ] \lambda6584/\mathrm{H}\alpha$, it provides a clear separation between star-forming galaxies, AGN (Seyfert and LINERs), and composite systems. Theoretical and empirical boundaries, such as the maximum starburst line from \citet{kewley2001theoretical}, the empirical demarcation of composite systems from \citet{kauffmann2003host}, and the LINER/Seyfert division from \citet{stasinska2006semi}—help interpret the nature of ionizing sources.

\begin{figure}[h!]
    \centering
    \includegraphics[width=1\linewidth]{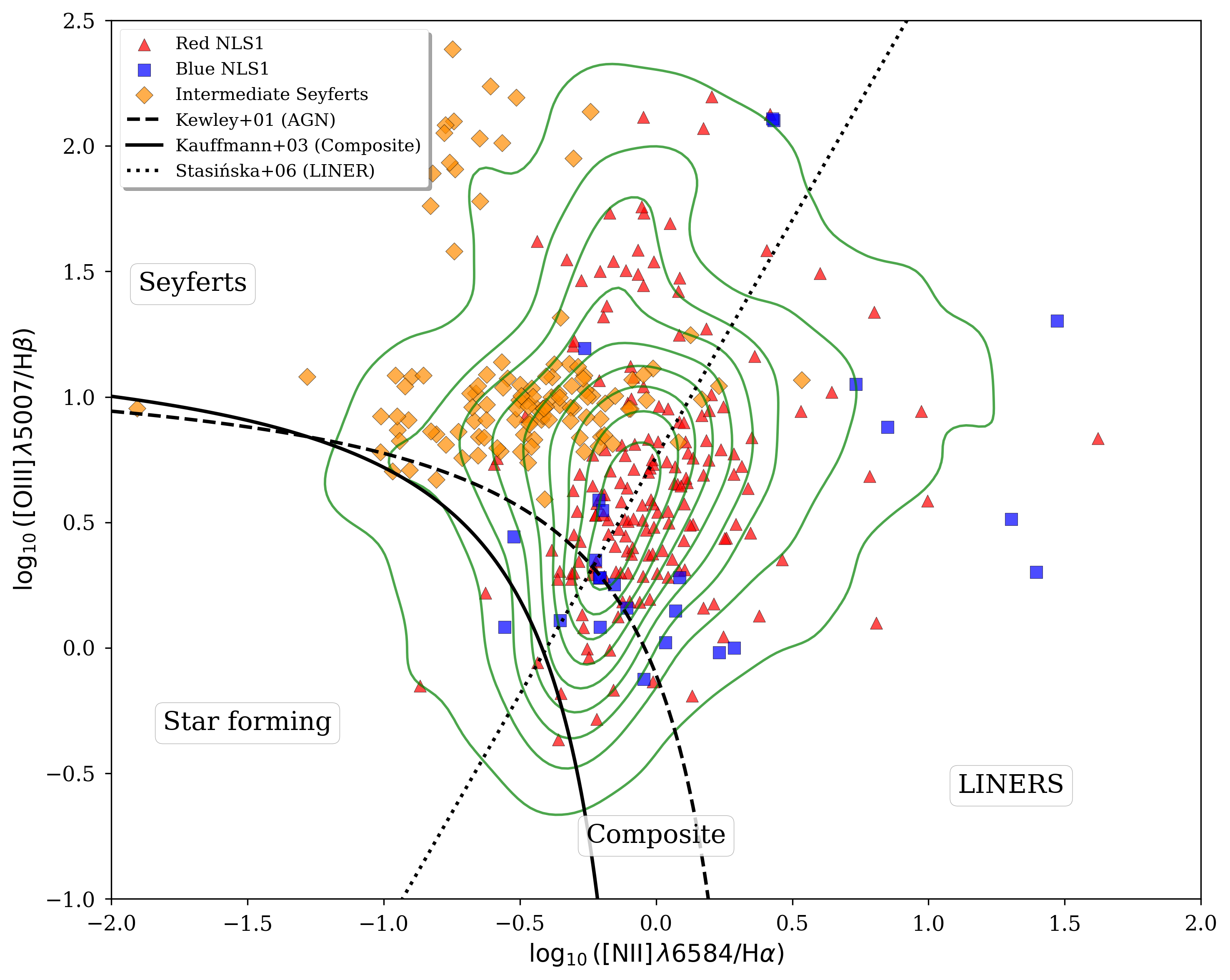}
    \caption{The BPT diagram for the all the three detected anomaly groups along with all the NLSy1 galaxies in the \protect\cite{paliya2024narrow} catalog (depicted by green contour lines). The respective regions of the BPT diagram are annotated.}
    \label{fig:BPT Diagram}
\end{figure}

In Fig.~\ref{fig:BPT Diagram}, we show the BPT distribution of our two anomaly classes of NLSy1 galaxies—red, blue NLSy1s and Intermediate Seyferts—overlaid on the classical AGN-star formation diagnostic lines. The overall distribution lies well above the star-forming region and nearly coincident on the \citet{stasinska2006semi} line, confirming their AGN nature. This supports their classification as Seyfert and/or NLSy1 galaxies, where high-ionization emission originates from the AGN’s NLR.

We do not have enough samples with emission line parameters for the \blues (their high redshift limits the coverage of \halpha line), due to which it would be impossible to make a comment about the nature of these objects using the BPT diagram. For the Intermediate Seyferts and red NLSy1s, a closer inspection reveals intriguing distinctions. The red NLSy1s, while also occupying the Seyfert  and composite regions, are mostly concentrated in the LINER region. They show a broader spread along the horizontal axis and appear slightly shifted toward lower $[ \mathrm{O\,III} ]/\mathrm{H}\beta$ ratios, suggesting a somewhat weaker ionizing continuum or enhanced Balmer emission (in case of star-forming galaxies), consistent with their observed lower AGN luminosity and host-dominated continua (as discussed in Sect.\ref{subsec:ebv}). \ldnls cluster completely in the Seyfert boundary and even extend towards the region with highest $[ \mathrm{O\,III} ]/\mathrm{H}\beta$ ratios. This shift toward elevated ratios along with the higher velocity dispersion (indicated by significantly higher \oiii EW; Fig.~\ref{fig:g-r_vs_OIII}) could indicate a stronger AGN ionizing flux \citep[see][]{ulivi2024feedback}. Their location suggests a higher incidence of either intermediate AGN activity with harder continuum or orientation effects leading to diluted ionizing radiation fields. The number of galaxies in each class is given in Table \ref{table:Anomalies per class}.

\begin{table}[h!]
\caption{Number of galaxies in each class of the BPT diagram.}
\label{table:Anomalies per class}
\centering
\begin{tabular}{|c | c | c | c | c|} 
 \hline
  Type & Star Forming & Composite & LINERS & Seyfert \\ [0.5ex] 
  \hline
    Red & 4 & 17 & 69 & 156 \\ 
    Blue & 1 & 5 & 12 & 4 \\ 
    Intermediate & 0 & 5 & 6 & 236 \\ 
   \hline
\end{tabular}
\end{table}

Together, the BPT analysis confirms the AGN identity of all three types while unveiling systematic differences in their ionization properties and central engine strength. These differences are likely tied to a combination of intrinsic AGN power, orientation, and possibly host galaxy conditions.

\begin{figure}[h!]
    \centering
    \includegraphics[width=1\linewidth]{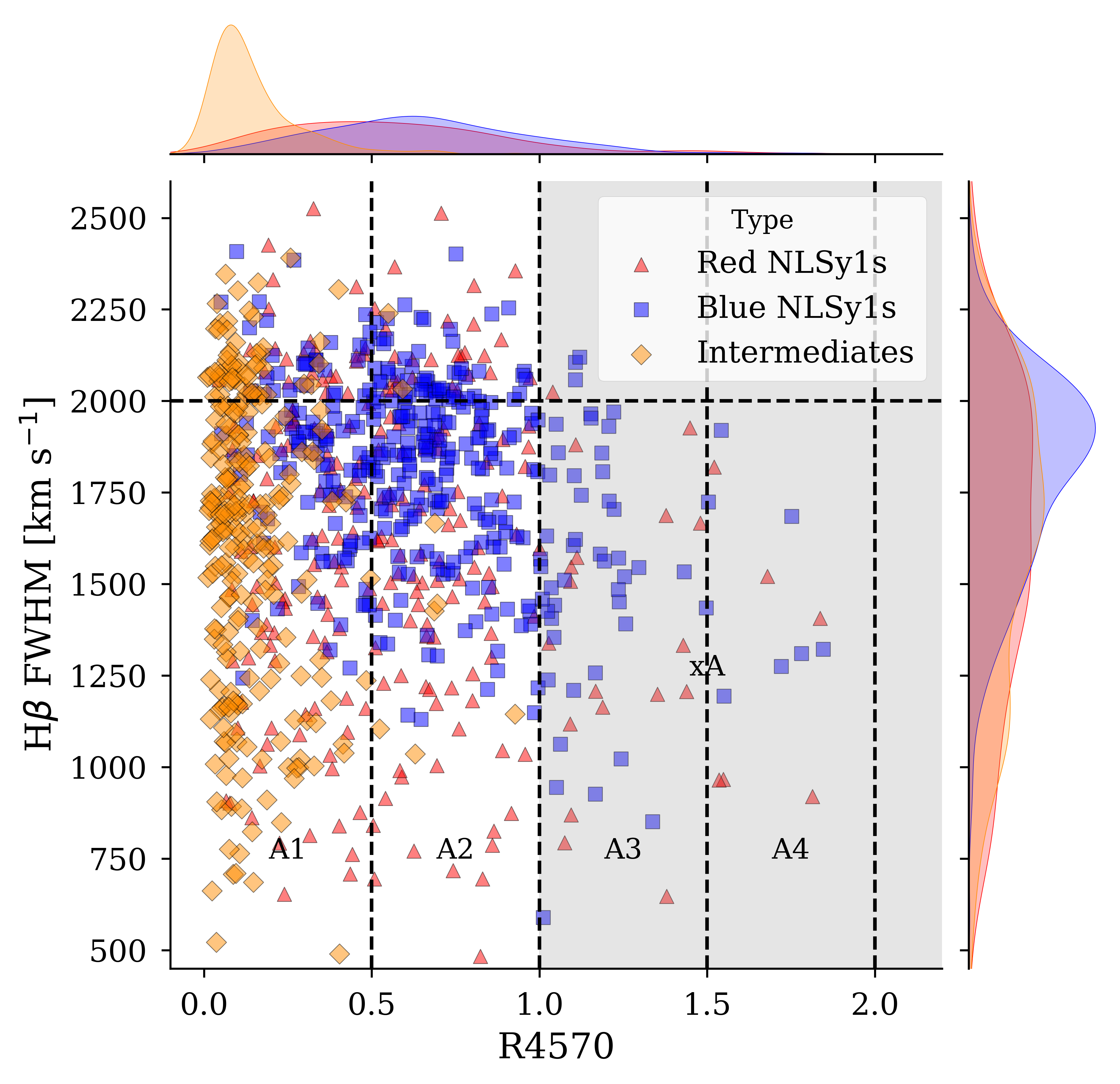}
    \caption{Eigenvector 1 diagram (H$\beta$ FWHM versus R4570) for red, blue NLSy1s and Intermediate Seyferts, represented as a scatter plot. The histograms on the top and right spline represent the distribution of R4570 and H$\beta$ FWHM respectively for all three classes. The gray region indicating \feii strength more than 1 (R4570 $>1$) represents the xA population.}
    \label{fig:EV1 diagram}
\end{figure}

\subsection{Eigenvector 1 diagram}
The quasar main sequence was formulated by \citet{marziani2001searching}, where they made the eigenvector 1 (EV1) diagram by plotting H$\beta$ FWHM versus \feii strength. They divided the quasar population into two major distinctions i.e. population A, quasars with H$\beta$ FWHM less than 4000 \kms and population B quasars with H$\beta$ FWHM greater than 4000 \kms. All NLSy1 galaxies by definition fit into population A as their H$\beta$ FWHM is defined to be less than 2000 \kms. Additionally, the population A is binned on the basis of \feii strength (R4570) into four subpopulations namely A1 ($0-0.5$), A2 ($0.5-1$), A3($1-1.5$) and A4 ($1.5-2$), with A4 being the strongest accretion sources and A1 being the weakest. A3 and A4 comprise the xA population indicating the \say{extreme} population A with strong \feii emission.

Fig.~\ref{fig:EV1 diagram} shows the EV1 diagram for the anomalous NLSy1 galaxies detected in this work. The \ldnls marked in yellow scatter points are present in the lowermost region of \feii strength (R4570), covering mostly the A1 region, indicating lowest accretion rate of the three classes. On the other hand, the \reds and \blues are majorly present in the A2 region and are spread all the way to A4 indicating high accretion rates evident by their strong \feii strength. As discussed, \blues are mostly \say{quasar} like with strong AGN activity, while the \reds can be considered as perfect candidates for a NLSy1 galaxy. This is because, \reds they can be thought of galaxies where the AGN activity has just began and is not yet strong enough to wash out the host galaxy signature in the spectra.
%####################################################################################
%####################################################################################
\section{Conclusion} \label{sec: Discussion}
\begin{center}

\begin{table*}[ht]
\centering
\setlength{\tabcolsep}{18pt}  % Adjust horizontal space between columns
\renewcommand{\arraystretch}{1}  % Adjust vertical space between rows
\caption{Sample table listing the anomalous NLSy1 galaxies identified in this work.}
\label{table:sample_fulltable}
\begin{tabular*}{\textwidth}{lccccl}

\hline
Name & RA  & Dec  & Redshift & r-band  & Anomaly \\ 
& (deg) & (deg) & (z) & magnitude &  type \\
\hline
\hline
SDSS J092438.88+560746.8 & 141.1620 & 56.1296 & 0.025 & 13.87 & red \\ 
SDSS J103922.32-003404.0 & 159.8430 & -0.5677 & 0.771 & 18.71 & blue  \\ 
SDSS J032723.13+004422.4 & 51.8463 & 0.7395 & 0.138 & 17.80 & Intermediate \\ \hline
\end{tabular*}
{\raggedright Note: A complete version of the table is available in machine-readable format. \par}
\end{table*}
\end{center}

We applied the unsupervised machine learning technique-\texttt{SQuAD} to the spectra of 22656 NLSy1 galaxies obtained from SDSS DR17 to identify two groups of anomalous NLSy1s and a population of Intermediate Seyferts (misclassified as NLSy1 in the parent sample \citep{paliya2024narrow}). Table.\ref{table:sample_fulltable} provides a sample list, including the SDSS name, right ascension, declination, redshift, r-band magnitude and their final classification. We then characterized the spectroscopic, photometric, and physical properties of these three populations— red, blue NLSy1s and Intermediate Seyferts by combining multi-wavelength and multi-approach diagnosis. Below, we summarize the key findings for each group and highlight the broader implications of this work.

%#######################################################################
\subsection{Red NLSy1 galaxies}
We conclude them to be host-dominated, low-luminosity young AGNs with a smaller black hole mass. They exhibit a steep red continuum with weak \feii emission and moderate \oiii, dominated by host galaxy starlight (evidenced by strong D4000 breaks and \caii absorption). The modest E(B-V) $\sim 0.54$ values rule out heavy nuclear obscuration. Reddening is instead attributed to host galaxy dilution of a faint AGN continuum. In the WISE color distributions, the \reds scatter into starburst/spiral regions in mid-IR color space, confirming host dominance over AGN emission. Hence physically, \reds likely represent low-accretion-rate NLSy1s where the AGN activity has just began and hence is intrinsically weak, allowing the host galaxy to dominate the optical spectrum. Their prevalence at low redshifts suggests selection biases against detecting them at higher-z due to sensitivity limits. It was noted that nearly all the \reds for which the SDSS image exhibited an extended identifiable source, featured a bar. This goes in accordance with the findings of \citet{Crenshaw_2003} who report that the NLSy1 fueling is bar driven.

\subsection{Blue NLSy1 galaxies}  
These are highly luminous NLSy1s with larger blackhole masses and high-Eddington ratio. The \blues exhibit a blue continuum with strong \feii (see bottom right panel of Fig.~\ref{fig: derived properties}) and weak \oiii, consistent with high-Eddington accretion. In the WISE color distribution plot, they are concentrated in the AGN region $(W1-W2 \geq 0.8)$, indicating hot dust emission and minimal host contamination. Their population exhibits maximum at $z \sim 0.8$, aligning with the cosmic peak of quasar activity, and rare at low-z due to declining AGN luminosity density function. Physically, they can be thought as classic high-$L/L_{Edd}$ NLSy1s, possibly undergoing rapid black hole growth. Their bluer colors and \feii strength suggest a radiatively efficient accretion flow with low dust extinction. As compared to a typical NLSy1 galaxy, they exhibit a much steeper blue continuum, higher \feii emission with mean R4570 = 0.66 as compared to a mean R4570 of 0.46 for the entire \citet{paliya2024narrow} catalog. Additionally, \blues have higher bolometric luminosities with a mean $L_{BOL}=45.51$ as compared to $L_{BOL}$ of 44.74 to 44.89 for a typical NLSy1 galaxy \citep{paliya2024narrow,jha2022comparative}. The \blues also exhibit a higher black hole mass with log $M_{BH} = 7.32\pm0.02 M_\odot$ as compared to $6.98\pm0.04$ for a typical NLSy1 \citep{rakshit2017catalog, rakshit2017optical,paliya2024narrow}. The \blues have a mean Eddington ratio of 1.13 as compared to a mean of 0.44 for the entire \citet{paliya2024narrow} catalog. The distributions of their absolute B band magnitude (Fig.~\ref{fig:MagB_Distribution}) and WISE colors (Fig.~\ref{fig:Wise Analysis}) suggest that these sources are more like a narrow-line quasar than an NLSy1 galaxy. 

\subsection{Intermediate Seyfert galaxies}

\begin{figure}[h!]
    \centering
    \includegraphics[width=1\linewidth]{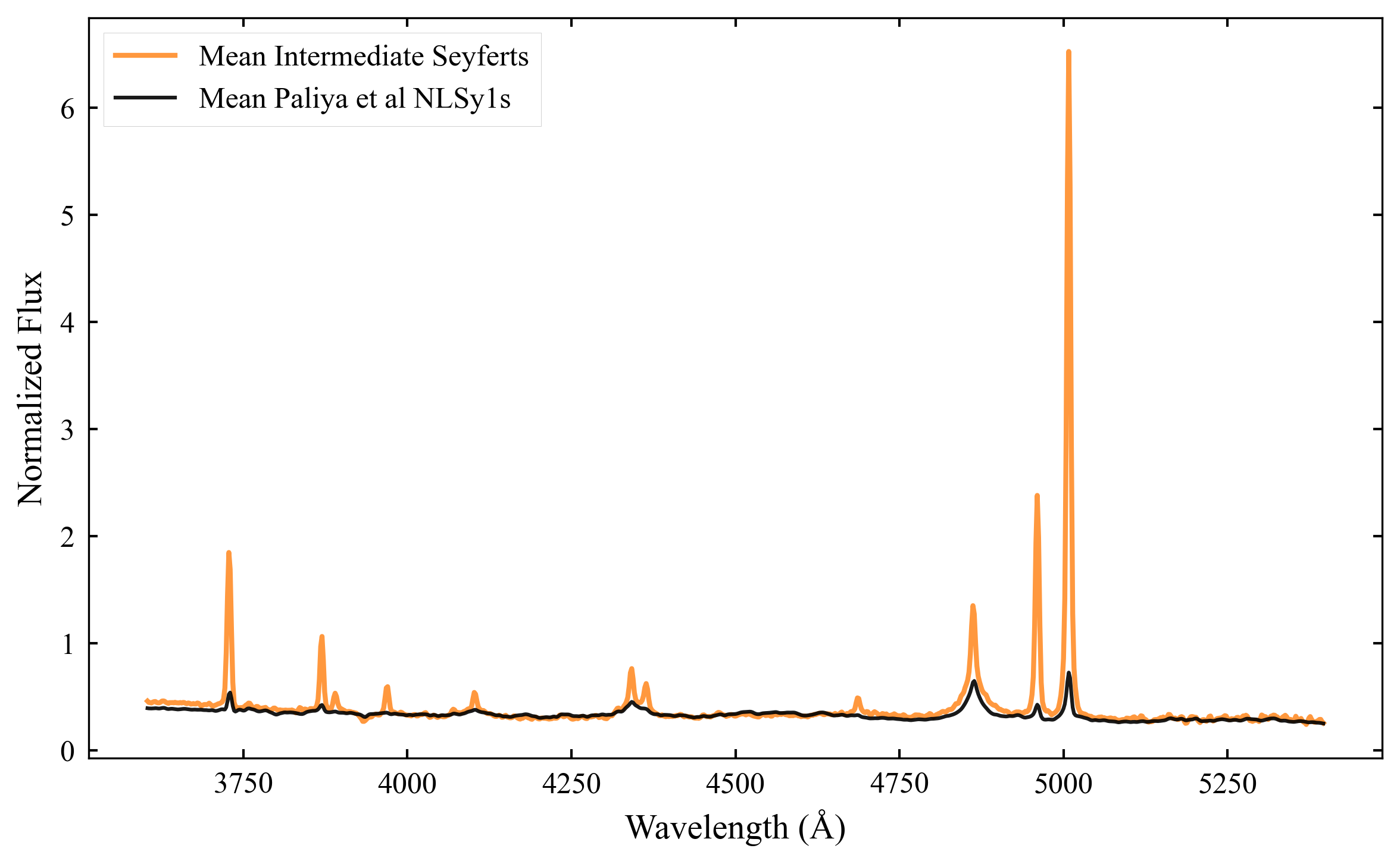}
    \caption{Mean composite spectrum for the Intermediate Seyferts (orange) as compared to the mean composite of all the NLSy1s in the \protect\citet{paliya2024narrow} catalog (black). The \protect\citet{paliya2024narrow} composite spectrum is scaled arbitrarily to match the continuum of the \ldnls for better comparison.}
    \label{fig:eld versus paliya total}
\end{figure}

This group is distinct from the NLSy1s and represents a different subset of Seyfert galaxies called Intermediate Seyferts. These are primarily identified by the presence of composite emission line profiles \citep{osterbrock1976ngc,cohen1983narrow}, i.e. a narrow component (characteristic of a Seyfert 2 galaxy) superimposed on a broad (typical of a Seyfert 1 galaxy). This group is distinguished by their spectra, which are almost entirely devoid of continuum emission and dominated by strong atomic emission lines (see Fig.~\ref{fig:eld versus paliya total}). Their spectral appearance resembles that of objects with an exposed NLR and highly suppressed or obscured accretion disk emission. As shown in Fig.~\ref{fig:g-r_vs_OIII}, these objects exhibit the highest \oiii equivalent widths among the anomaly types—an indicator often associated with edge-on viewing angles or optically thick circumnuclear material, as suggested by \citet{10.1111/j.1365-2966.2010.17843.x}. This interpretation aligns with the lack of observable continuum emission, supporting the idea that these sources may be obscured by orientation or structural effects. Moreover, the \ldnls display elevated values of the $W1-W2$ infrared color, likely due to thermal dust re-emission, consistent with their inferred intermediate inclination geometry. However, their $E(B-V)$ distribution suggests little to no effect of dust reddening in their optical spectra, contradicting the edge-on geometric interpretation. Their positions on the WISE color-color diagram (see bottom panel of Fig.~\ref{fig:Wise Analysis}) span both AGN and starburst-composite regions, implying a diversity in dust geometries or evolutionary phases. These combined features suggest that the \ldnls may represent a transitional or structurally unique phase of AGN activity. Overall, the physical properties of \ldnls \textemdash as suggested by our analysis is, in general, contrasting to those of a typical NLSy1 galaxy. The \ldnls, represent a contaminant class in the \citet{paliya2024narrow} catalog, wrongfully labeled as NLSy1 galaxy and instead are a class of Intermediate Seyferts. This interpretation is consistent with the detailed reanalysis of SDSS J21185.96-073227.5 by \citet{2020A&A...636L..12J}, who revised its classification from a $\gamma$-ray emitting NLSy1 to an intermediate Seyfert.

The properties of \ldnls such as hard-ionization continua, narrow and strong emission lines, virtual lack of continuum is remarkably similar to the \say{\civ peakers} presented by \cite{tiwarisquad1}. These objects exhibit rare emission lines such as {Ne\,{\sc v}$\lambda3345$}, {[Ne\,{\sc v}]$\lambda3427$}, {[Ne\,{\sc vi}]$\lambda3869$} \citep{cohen1983narrow}, that are absent in a typical quasar or NLSy1, providing us with an opportunity to explore new areas of quasar astrophysics. These high ionization lines have been used to determine physical parameters such as the inclination of the sources \citep{2025A&A...697A.175S}, \citet{Trakhtenbrot_2025} use the {[Ne\,{\sc v}]$\lambda3426$} line to calculate appropriate scaling to calculate AGN luminosity and SMBH mass in high redshift AGNs. They also use the ratios of these high ionization lines to run photo ionization modeling, which were then used to plot BPT diagrams curated for high redshift sources. \cite{MatthewTemplePaper} have provided a framework to study the co-evolution of \civ and \heii emission lines and the gas, black hole and accretion properties. We plan to use this framework in an upcoming work for the analysis of these  sources. 

By integrating spectral decomposition, dust extinction analysis, mid-IR diagnostics, and emission-line kinematics, this work provides a unified framework for understanding NLSy1 and Seyfert diversity. The anomalies identified here—whether host-dominated, accretion-driven, or geometrically obscured—highlight the complex interplay between AGN physics and host galaxy environments. Future surveys (e.g., Large Synoptic Survey Telescope (LSST), SDSS-V) will expand this taxonomy, probing fainter and rarer AGN subpopulations to refine our picture of black hole growth and feedback. Additionally, given the distinct and extreme properties of Intermediate Seyferts, we are conducting a deeper spectroscopic and multi-wavelength studies of these intriguing objects to further constrain their physical nature and evolutionary context.

\begin{acknowledgements}

We thank the anonymous referee for the feedback which has significantly helped to improve the paper. AT acknowledges and thanks the Indian Institute of Astrophysics (IIA) for the financial and infrastructural support provided, and to Aarya H for helping with the PyQSOFit fittings of the red NLSy1s. MV acknowledges support from Department of Science and Technology, India - Science and Engineering Research Board (DST-SERB) in the form of a core research grant (CRG/2022/007884).
Funding for the Sloan Digital Sky Survey IV has been provided by the Alfred P. Sloan Foundation, the U.S. Department of Energy Office of Science, and the Participating Institutions. SDSS acknowledges support and resources from the Center for High-Performance Computing at the University of Utah. The SDSS web site is www.sdss4.org.

SDSS is managed by the Astrophysical Research Consortium for the Participating Institutions of the SDSS Collaboration including the Brazilian Participation Group, the Carnegie Institution for Science, Carnegie Mellon University, Center for Astrophysics | Harvard \& Smithsonian (CfA), the Chilean Participation Group, the French Participation Group, Instituto de Astrofísica de Canarias, The Johns Hopkins University, Kavli Institute for the Physics and Mathematics of the Universe (IPMU) / University of Tokyo, the Korean Participation Group, Lawrence Berkeley National Laboratory, Leibniz Institut für Astrophysik Potsdam (AIP), Max-Planck-Institut für Astronomie (MPIA Heidelberg), Max-Planck-Institut für Astrophysik (MPA Garching), Max-Planck-Institut für Extraterrestrische Physik (MPE), National Astronomical Observatories of China, New Mexico State University, New York University, University of Notre Dame, Observatório Nacional / MCTI, The Ohio State University, Pennsylvania State University, Shanghai Astronomical Observatory, United Kingdom Participation Group, Universidad Nacional Autónoma de México, University of Arizona, University of Colorado Boulder, University of Oxford, University of Portsmouth, University of Utah, University of Virginia, University of Washington, University of Wisconsin, Vanderbilt University, and Yale University.
\end{acknowledgements}

\bibliography{SQuAD_NLSy1}{}
\bibliographystyle{aasjournalv7}

\end{document}